\documentclass[12pt]{article}

\usepackage{amsmath,amssymb,graphicx} 
\usepackage{epsf}
\usepackage{pstricks}

\newcommand{\beq}{\begin{eqnarray}}
\newcommand{\eeq}{\end{eqnarray}}

\newcommand{\centeron}[2]{{\setbox0=\hbox{#1}\setbox1=\hbox{#2}\ifdim

\wd1>\wd0\kern.5\wd1\kern-.5\wd0\fi \copy0

\kern-.5\wd0\kern-.5\wd1\copy1\ifdim\wd0>\wd1
                                          \kern.5\wd0\kern-.5\wd1\fi}}
\newcommand{\ltap}{\>\centeron{\raise.35ex\hbox{$<$}}
                                  {\lower.65ex\hbox{$\sim$}}\>}
\newcommand{\gtap}{\>\centeron{\raise.35ex\hbox{$>$}}
                                  {\lower.65ex\hbox{$\sim$}}\>}

\newcommand\ZZ{\hbox{\zfont Z\kern-.4emZ}}
\font\zfont = cmss10 

\def\gappeq{\mathrel{ \rlap{\raise.5ex\hbox{$>$}}
                         {\lower.5ex\hbox{$\sim$}}  } }
\def\lappeq{\mathrel{ \rlap{\raise.5ex\hbox{$<$}}
                         {\lower.5ex\hbox{$\sim$}}  } }
\newcommand{\mpl}{M_{Pl}}
\newcommand{\reta}{R_{\eta}}
\begin{document}
\begin{titlepage}

\vskip.5cm
\begin{center}
{\LARGE   Black Holes and Quantum Gravity at the LHC
    \vspace{.2cm}}

\vskip.1cm
\end{center}
\vskip0.2cm

\begin{center}
{\bf Patrick Meade and Lisa Randall}
\end{center}
\vskip 8pt

\begin{center}
{\it Jefferson Physical Laboratory\\
Harvard University\\
Cambridge, MA 02138, USA } \\
\vspace*{0.3cm} {\tt
meade@physics.harvard.edu,randall@physics.harvard.edu}
\end{center}

\vglue 0.3truecm

\begin{abstract}
\vskip 3pt \noindent
\end{abstract}
We argue that the highly studied black hole signatures based on
thermal multiparticle final states are very unlikely and only
occur in a very limited parameter regime if at all. However, we
show that if the higher-dimensional quantum gravity scale is low,
it should be possible to study quantum gravity in the context of
higher dimensions through detailed compositeness-type searches.

\end{titlepage}

\newpage

\renewcommand{\thefootnote}{(\arabic{footnote})}
\section{Introduction}
\setcounter{equation}{0} \setcounter{footnote}{0} One of the most
exciting possibilities for the LHC is the discovery of small
higher-dimensional \cite{extrad} black holes that could be formed
when two sufficiently energetic particles collide
\cite{Dimopoulos:2001hw,Giddings:2001bu,Argyres:1998qn,banksfischler}.
Ideally, such black holes would decay isotropically to many
energetic particles, in keeping with the prediction of thermal
Hawking radiation~\cite{Hawking:1974sw}. However, over most of the
viable parameter space, this expectation is not very realistic.
Once inelasticity and black hole entropy are accounted for, it is
clear that multiparticle final states are very suppressed, since
only black holes produced well above threshold have sufficient
entropy. The falling parton distribution functions(PDFs) more than
compensate for the rise in black hole production with energy so
most strong gravity events will occur at the lowest possible
energy scale.

Nonetheless, all is not lost. Even when the energy is too low to
produce truly thermal black holes, which require sufficiently high
entropy and energy, we would nevertheless expect signs of quantum
gravity if higher dimensional gravity gets strong at a scale not
too far above a TeV. Strong gravity is likely to result in more
spherical final states, even for those final states with low
multiplicity, which would therefore be measured as much more
transverse than background.  As we will show, over most regions of
expected parameter space for higher dimensional models, we expect
a significant change in the rate of highly transverse two particle
final states to occur at the quantum gravity scale, both jet-like
and leptonic, although the latter rate which is smaller spans a
smaller region of parameter space. Strong gravity should be
testable through standard compositeness tests.

In fact, the threshold for a rise in the 2$\rightarrow$2
scattering cross section is almost  inevitably lower than the
black hole production threshold. Though not necessarily a true
thermal black hole, these final states, if they occur, will
nonetheless tell us about quantum gravity. In fact, in the
thermal regime, black holes wouldn't give us {\em any} insight
into quantum gravity (except to confirm existing theoretical
predictions). In the region at or below the true thermal black
hole threshold, assuming strong gravity effects don't turn on or
off suddenly at the black hole scale, we could in principle learn
a lot by studying the two particle final states, in particular the
angular distribution and the energy dependence of the angular
distribution which would truly be quantum gravity results, not
interpretable in terms of  a classical calculation.

Furthermore we will see that there is sufficient information to
distinguish not only black hole type effects, but different forms
of string amplitudes. This can in principle probe the effects of
curvature or non-string objects in the theory as well.  Moreover, we don't
expect only strong gravity effects if higher dimensional theories
are right. We should in that case find indications of KK final
states at lower energy. In that case there would be indications
whether composite-type effects might be associated with quantum
gravity to help us disentangle it from other strongly interacting
physics. In what follows, we will see other possible distinctive
features of gravitational physics that might help distinguish
among possibilities.  Thus what we are saying is that even
existing compositeness searches don't just tell about strong gauge
dynamics-they could in principle tell us about gravity as well.
 We show how we can hope to learn about black hole
production and quantum gravity by studying the energy dependence
of the high $p_T$ dijet or leptonic cross section. We consider the
implications of  a rise or fall in the cross section and what the
energy dependence might teach us about quantum gravity.

We stress that although the two particle final state signal is
unlikely to probe thermal black holes in the accessible energy
range, it is of great interest as a way of probing quantum
gravity. The rate as a function of energy as well as the angular
distribution can differ significantly in various scenarios of
quantum gravity. Furthermore in almost any scenario we expect the
two particle final state to demonstrate effects of quantum gravity
well before the proposed multiparticle final states characteristic
of thermal black holes. Furthermore whereas we know the
predictions for the semiclassical regime, independent of the
particular theory of quantum gravity, the threshold regime can
potentially distinguish among them.

Others have considered the effects of specific gravitational
effects on higher-dimensional operators and how they can be
constrained by existing searches. Ref. \cite{hewett} considered a
dimension-8 operator, Ref. \cite{giudice} considered graviton
loops generating a dimension-6 operator, Ref. \cite{han1,peskin}
considered string-generated dimension-8 operators and string
resonances, Ref. \cite{antoniadis} considered dimension-6
operators from string theory. Our point is to view compositeness
searches more generally and to learn how to distinguish among the
possibilities rather than to constrain the scale of any one
particular model. Furthermore we emphasize that the gap between
the quantum gravity scale and the true black hole threshold should
be a good source of deviations in 2$\rightarrow$2 scattering and
probably yields a much better reach and more insight than
multiparticle searches.

\section{Black Hole Production and Decay}
The large black hole cross section estimate stems from the
classical cross section that is proportional to the geometrical
area set by the Schwarzschild radius $r_S$:
\begin{equation}\label{roughxs}
\sigma(E)\sim \pi r_S(E)^2.
\end{equation}
This geometrical cross section implies
\begin{equation}
\sigma(E)\sim \frac{1}{M^2} \left(\frac{E}{M}\right)^\alpha
\end{equation}
where $M$ is the effective scale of quantum gravity and
$\alpha\leq 1$ for higher-dimensional black holes.  Thus for
instance at the LHC one might expect a parton-parton cross section of
size at least $\sim \frac{1}{M^2}$, which for $M\sim 1$ TeV
corresponds to an enormous rate of approximately 100 pb which for
100 $fb^{-1}$ luminosity would yield ten million events.  The
basic reason why this cross section is so large compared to the
production of a particle with TeV mass in a typical beyond the SM
theory is the lack of any small couplings, such as gauge couplings
in the cross section and the absence of phase space suppression
factors.  However, this estimate ignores several major
considerations and uncertainties in the black hole
production\cite{Giddings:2001bu,Dimopoulos:2001hw} and decay cross
sections that we discuss in the rest of this section.

There have been relatively few studies of the phenomenological
consequences of RS black holes, and thus in addition to
elaborating the points above we will also expand further upon this
case throughout the paper and in Appendix~\ref{app:rsbh}.
Landsberg \cite{landsbergtalk} discussed RS black hole signatures,
but used more optimistic assumptions for parameter space than are
now experimentally allowed and neglected the inelasticity that we
will soon discuss. Ref. \cite{Anchordoqui:2002fc} considered black
holes that might arise in warped five-dimensional space in the
context of cosmic ray searches.  For further references see
Appendix~\ref{app:rsbh}. We will see in Appendix~\ref{app:rsbh}
that in the energy range between $\tilde{M}$ and $(M/k)^2
\tilde{M}$, where $M$ is the five dimensional Planck
scale($\tilde{M}$ is $M$ reduced by a warp factor) and $k$ is
related to the AdS curvature, we expect to a good approximation
conventional five-dimensional black holes. Of course, in the RS
case where approximately  flat space black holes occur only over a
limited energy range,  we would need $M/k$ large enough to permit
high entropy black holes.

\subsection{Criteria for Black Holes}\label{criteria}

\begin{figure}[h]
\begin{center}
\includegraphics[width=10cm]{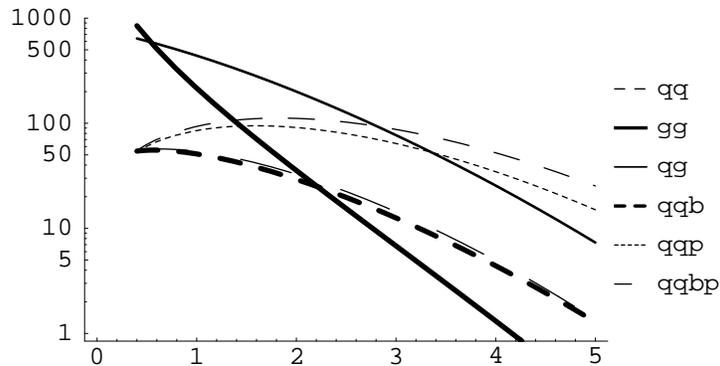}
\end{center}
\caption{Arbitrarily normalized parton-parton luminosity plot as a
function of $\sqrt{\hat{s}}$ to show the relative contributions of
initial state partons.} \label{pdfplot}
\end{figure}

The production cross section in (\ref{roughxs}) depends only on
the mass scales involved and thus appears to be a very simple
quantity to understand. Unfortunately however there are
ambiguities associated with both of the two scales in the problem,
$M$ and $M_{BH}$. Since one makes rough estimates assuming black
holes start forming at a scale $M$, and due to the falling
PDFs\footnote{The effective scaling of the PDFs can be summarized
in terms of a parton luminosity.  See for instance Figure 69
of~\cite{Campbell:2006wx}.  The drop in the parton luminosity at
the LHC depends on the mass range of interest, for instance for
$qq$ and $\sqrt{\hat{s}}\sim 1-2$ TeV the luminosity
  drops off approximately as $\sim 1/\hat{s}$ while for higher
invariant mass it can drop off as $\sim 1/\hat{s}^{4}$.} the rate
changes dramatically depending on the scale at which black holes
start to form, it is critical to keep track of the different
conventions for the Planck scale and the relationships among them
so that we can unambiguously compare rate predictions. See Figure
\ref{pdfplot} to see the different relative contributions to
$2\rightarrow 2$ scattering from the pdfs. These will be helpful
in understanding results throughout the paper.

Different authors have used different conventions for normalizing
the Planck scale. We define $G_D$ with the Myers-Perry
convention~\cite{myersperry}
\begin{equation}\label{mpln}
\frac{1}{16\pi G_D}\int d^{D+1}x \sqrt{g} R
\end{equation}
and define $\mathcal{L}_N$ as the normalization of the
Einstein-Hilbert action for which (\ref{mpln}) gives $1/16\pi
G_D$.  In the case of $n$ extra dimensions, the PDG convention
\cite{giudiceconvention} is
  $\mathcal{L}_N = {M_D^{n+2}/2} (2\pi)^n$ whereas the early analysis of Dimopoulos and
Landsberg \cite{Dimopoulos:2001hw} used $M_P^{n+2}/16 \pi$.
Although neither analysis was done for case of one extra dimension
due to the constraints on $n=1$ ADD type set ups~\cite{pdg}, there
is a range of mass scales for which approximate five-dimensional
flat space black holes would be the most appropriate description
for RS models (see Appendix~\ref{app:rsbh}).  To illustrate the
convention dependencies we give their formulae for $n=1$ so as to
compare to RS, in which case their formulae reduce to
$\tilde{M}_P^3/16 \pi$ and  $\tilde{M}_D^3/4 \pi$, which should be
compared to $\tilde{M}^3/2$, which is the RS convention, where the
tilde indicates the warped version of the various Planck scales.
Although just conventions, it is important to bear these
conventions in mind when interpreting results.

The Schwarzschild radius of the black hole given
in~\cite{myersperry} for the $(4+n)$-dimensional case is
\begin{equation}
r_S=\left(\frac{M_{BH}\Gamma\left(\frac{n+3}{2}\right)}{\mathcal{L}_N
(n+2) 2\pi^\frac{n+3}{2}} \right)^{\frac{1}{n+1}}
\end{equation}
where the scale is understood to be appropriately warped in the RS
case (for details see the Appendix), which reduces to
\begin{equation}
\left(\frac{M_{BH}}{\mathcal{L}_N 6\pi^2}\right)^{1/2}
\end{equation}
for the case of $n=1$. Using the RS normalization of the action we
find that the Schwarzschild radius in RS1 is given by
\begin{equation}
r_S=\left(\frac{M_{BH}}{\tilde{M}^3 3\pi^2}\right)^{1/2}.
\end{equation}
For the case of one extra dimensions, the DL and PDG conventions
would give
\begin{equation}
r_S^{DL}=\left(\frac{8 M_{BH}}{M_P^3
3\pi}\right)^{1/2}\;\;\;r_S^{pdg}=\left(\frac{2 M_{BH}}{M_D^3
3\pi}\right)^{1/2}
\end{equation}
where $M_P$ and $M_D$ are the higher-dimensional Planck scales in
the two cases.

Although just a convention, the numerical relationships mean that
if we take $r_S\sim 1/M$ as the threshold for black hole
production,   comparing the two formulations of the Schwarzschild
radius in the case of~\cite{Dimopoulos:2001hw} we would find that
black holes would be produced at energies $\sim M_{P}$, while
in~\cite{fenginelast} black holes would be produced at a scale of
$\sim 4^{1/3}\sim 1.6 M_D$ while the  convention would yield $
(8\pi)^{1/3}\tilde{M} \sim 2.9 \tilde{M}$. These conventions are
clearly significant in interpreting the meaning of the black hole
energy reach for the LHC and comparing to experimental
constraints. Of course the physical answers are not convention
dependent. When we compare the scales relative to threshold
production to the current experimental bounds on KK masses, the
convention dependence drops out.

The real question is the black hole threshold where black holes
start to form.  Of course at center of mass energies much greater
than the higher-dimensional Planck scale, $M$, we know black holes
will be produced. However, the precise threshold is ambiguous. $M$
is after all convention dependent. Though we will assume $E>M$ is
necessary, it is clearly not sufficient.

Since we don't know the precise threshold for a truly thermal
black hole, it is useful to define a parameter $x_{min}$ that
tells how far above the relevant Planck scale the semiclassical
prediction applies \cite{Giddings:2001bu}. This could be defined
relative to an  arbitrary threshold mass or relative to the
convention-dependent Planck scale. We will use the latter with the
understanding that $x_{min}$ is unknown either way and is simply a
parameter.  In our analysis we will give results as a function of
$M$ and $x_{min}$. We consider criteria for $x_{min}$ below. Note
that we would want $x_{min}$ for RS to be less than $(M/k)^2$
where the curvature becomes relevant as outlined in
Appendix~\ref{app:rsbh}.

Keep in mind that in addition to significantly reducing the black
hole production cross section,  the existence of a nontrivial
$x_{min}$   obscures our ability to extract fundamental parameters
from the black hole cross section. The overall cross section
depends very strongly on $x_{min}$ since as we have already noted,
the rapid fall-off of the PDFs makes us very sensitive to the mass
threshold where black hole production can begin. This means that
any potential bounds from an LHC experiment on black hole
production rates  is only indirectly related  to the fundamental
scale of quantum gravity.  For instance if one finds an excess of
events attributed to black hole it is unclear  how  to translate
back to the scale $M$ involved if one is only looking on the tail
of a distribution.

Without knowing more about the threshold behavior of black hole
production, the dependence of the cross section on the fundamental
Planck scale is insufficient to extract that parameter, which can
be mimicked by an alternate choice of $x_{min}$. In principle, the
energy-dependence of the cross section can  be useful in
extracting the number of dimensions (if we know the PDFs
sufficiently accurately), although in practice this will be very
challenging. In any case, this slope  won't determine the
higher-dimensional Planck scale. In principle, the differential
cross section can be used to extract the Planck scale since, once
it has turned on, the cross section depends on black hole mass
(not $x_{min}$). But without the energy-dependent inelasticity
factor (see below) this will be impossible. Furthermore,
uncertainties in PDFs and the experimental determination of energy
scale will also make this unlikely.

\subsection{Thermality}

Although difficult to quantify precisely, we now consider several
possible criteria for the formation of a truly thermal black hole.
Though not sufficient, we expect these to be some minimum
necessary criteria that will give some sense of  what $x_{min}$
should be.

The first criterion one might apply is that the Compton wavelength
of the colliding particle of energy $E/2$ lies within the
Schwarzschild radius for a black hole of given energy $E$. If we
define the threshold as the point where a wave with wavelength $4
\pi/E$ lies within the Schwarzschild radius for a black hole of
mass $E$, we find for ADD $n=6$ black holes this yields $x_{min}>
4.1$ (in the $M_D$ convention). Had we simply required $r_S>1/E$,
we would have the weaker criterion $x_{min}>0.44$. In the RS case,
we find with the stronger criterion that $x>16$, whereas with the
weaker criterion it should be greater than about 3.  We see that
this criterion in and of itself if fairly strong, and already will
make black hole production very small or {\em nonexistent} given
LHC parameters.

Even so, the above criterion is not necessarily sufficient  to guarantee a
black hole since we don't expect the semiclassical formula to
apply at the threshold determined above. For a black hole to be
truly thermal, we expect higher entropy is required and therefore
the threshold will be above the energy we just considered. There
are several additional criteria that we would want to be
satisfied, all roughly amounting to the fact that the black hole
should be sizable enough that the entropy is large. Although for
sufficiently large black holes, any criteria of the sort below
will be amply satisfied, as we have emphasized, the falling PDFs
tell us production is dominated by near-threshold objects.

For the criteria below, the following formula will prove useful.
For $n$ extra dimensions we have
\begin{equation}\label{rtrelation}
r_S=\frac{1+n}{4\pi T}=\frac{k(n)}{M_D} \left ( \frac{M_{BH}}{M_D}
\right)^{\frac{1}{1+n}},
\end{equation}
where
\begin{equation}
k(n)=\left(2^n \pi^{\frac{n-3}{2}}
\frac{\Gamma\left(\frac{n+3}{2}\right)}{2+n}\right)^\frac{1}{1+n}
\end{equation}

\begin{equation}\label{entropy}
S=\frac{1+n}{2+n} \frac{M_{BH}}{T_{BH}}
\end{equation}

It is also useful to consider the average number of particles
assuming the decay is mostly on the brane~\cite{braneradiate}. The
prediction for black hole decays in experiments have been
approached in a couple of ways, including  treating the decay as
instantaneous\cite{Dimopoulos:2001hw}, evolving with
mass\cite{fenginelast,Cavaglia:2003hg}, and sometimes including
the appropriate grey body factors for the extra-dimensional black
holes as well. These distinctions have an order one impact on the
average number of particles comparing for instance to an
instantaneous decay calculation with
\begin{equation}
\langle N \rangle= \left(\frac{ 2 \sqrt{\pi}}{n+1}\right)
\left(\frac{M_{BH}}{M_P}\right)^{(n+2)/(n+1)}\left(\frac{8
\Gamma((n+3)/2)} {2+n}\right)^{1/(n+1)}
\end{equation}
compared to one that evolved the black hole with mass and included
greybody factors
\begin{equation}
\langle N \rangle= \rho S_0=\rho \left(\frac{4\pi
k(n)}{2+n}\right)^{(n+2)/(n+1)}
\left(\frac{M_{BH}}{M_D}\right)^{(n+2)/(n+1)}.
\end{equation}
Allowing for the difference in the definitions of the Planck
scales, the instantaneous decay gives particle number a factor of
1.44 times that calculated by decaying over time.  The mass
scaling is in accordance with the mass-dependence of the entropy.

For the specific cases we will be interested we list the average
number of particles emitted for ADD $n=6$
\begin{equation}\label{decay1}
\langle N \rangle\sim \frac{4\pi \rho
k(6)}{8}\left(\frac{M_{BH}}{M_D}\right)^{\frac{8}{7}}
\end{equation}
with
\begin{equation}\label{decay2}
\rho=\frac{\sum c_i g_i \Gamma_i\zeta(3)\Gamma(3)}{\sum c_i f_i
\Phi_i\zeta(4)\Gamma(4)}
\end{equation}
which defines a ratio of multiplicities and greybody factors
defined in~\cite{fenginelast}. For RS $n=1$\footnote{We
approximate the greybody factor for $n=1$ as the same as that for
$n=6$} we find
\begin{equation}
\langle N \rangle\sim \frac{4\rho}{3\sqrt{3}}
\left(\frac{M_{BH}}{\tilde{M}}\right)^\frac{3}{2}
\end{equation}
Notice that $\langle N\rangle =\rho S$.

In what follows below, we will use the grey-body corrected
time-dependent decay estimate. Of course, near threshold, all
these formulae are unreliable  but give an idea of what one might
expect.

\begin{itemize}
\item Preskill et al \cite{Preskill:1991tb} give the criterion
$\vert\partial T/
\partial M \vert <<1$, which is equivalent to the change in Hawking
temperature per particle emission should be small. This condition
is equivalent to the entropy~(\ref{entropy}) being large.  More
specifically, $\partial T /\partial M\sim 1/((n+2)S)$. The
improvement of this bound scales as $x_{min}^{\frac{2+n}{1+n}}$.
This is not as strong a constraint as the other criteria we give
below, for RS and ADD the constraint is satisfied already in both
cases for $x_{min}=1$

   \item  We would also want the energy of any individual
degree of freedom in a thermal bath to be much less than the black
hole mass. This gives the criterion $(n+3) T <M$ or equivalently
$dM/dN<<M$--that is, any one individual degree of freedom should
not carry a significant fraction of the energy.  This particular
criterion is satisfied in ADD and RS for $x_{min}\gtrsim 2$. This
condition is slightly subtle in the case when brane black hole
decays are allowed, since the energy per degree of freedom for
modes on the brane is reduced by roughly a factor of $3/(3+n)$
since brane modes can oscillate only along the brane directions.

You can also see this directly from the formulae for the rate of
change of energy and number of particles when decaying into
thermal d-dimensional states. In the former case, the decay rate
is determined by $\int d^d k f(E) E$, whereas in the latter case
it is determined by $\int d^d k f(E)$, where $f(E)$ is the
Boltzmann factor. The resulting ratio whose inverse determines
particle number is proportional to $d \xi(d+1)/\xi(d)$, which is
approximately $d$. That is, for decays into more dimensions
(fixing $T$), we have fewer particles since each particle carries
more energy. Even if the bulk modes don't dominate the decay, we
would still not want any single bulk mode to carry a significant
fraction of the black hole energy if we are to interpret the
decaying object as a higher-dimensional black hole.

This is a stricter criterion than above. We find one bulk degree
of freedom carries almost all the energy when $M_{BH}\sim 3
\tilde{M}$ in the case of RS, and slightly exceeds it in the case
of ADD $M_{BH}\sim 2 M_D$($n=6$). Clearly we would want $M_{BH}>M$
in both cases as the bound improves as $x_{min}^{(2+n)/(1+n)}$,
again scaling as the entropy.

Of course we should keep in mind this is the criterion for {\it
one} degree of freedom in the bulk to carry all the mass. Clearly
for a thermal black hole, we would want many particles carrying
the energy, so the bound would be much stronger. For example, the
maximum experimental reach on $x_{min}$ for ADD $n=6$ is about 6,
which would correspond to only 3 bulk particles! For RS, the
maximum $x_{min}$ is about 10, corresponding to at most about 5 or
6 particles sharing the energy, which also seems inadequate for a
truly thermal state.

   \item  We want the black hole lifetime to be bigger than
$1/M$, so that the black hole appears as a
resonance~\cite{Giddings:2001bu}. This criterion scales roughly as
the number of degrees of freedom modified by grey-body factors.
This is borderline for $n=6$ and reasonably well satisfied for
$n=1$.   For completeness we give the formula for the lifetime in
ADD:
\begin{eqnarray}\label{lifetime}
\tau &=& \frac{(4 \pi)^4 k(n)^2 M_D^\frac{-2(2+n)}{1+n}
M_{BH}^\frac{3+n}{1+n}}{\alpha (1+n)^3 (3+n)}\\
\alpha &=& \frac{1}{2\pi} \left(\sum c_i f_i \Phi_i\right)
\zeta(4)\Gamma(4)
\end{eqnarray}
where the factors in $\alpha$ are defined in~\cite{fenginelast},
and correspond to multiplicities and greybody factors.  For the
specific case of $n=6$ we find
\begin{equation}
\tau = .7 \frac{x_{min}^{9/7}}{M_D}.
\end{equation}
In RS we can find $\langle N\rangle$ from (\ref{lifetime}) by
substituting $n=1$, and replacing $M_D$ with $\tilde{M}$ and
$k(1)$ with $1/3\pi^2$ to account for the RS normalization. The
result of this is that in RS the lifetime is given by
\begin{equation}
\tau = .38 \frac{x_{min}^{2}}{\tilde{M}}.
\end{equation}
Using these criteria we find that in ADD the criteria is satisfied
for $x_{min}\sim 1.3$ and in RS for $x_{min}\sim 1.6$.

   \item A sometimes stricter criterion in the case of black holes
that can decay on the brane is that the lifetime should exceed the
black hole radius, so that the black hole can reequilibrate as the
black hole decays primarily along the brane. This requires in the
ADD case that $x \gtrsim 3$ while for RS the constraint is
satisfied for any $x_{min}$.

   \item  The black hole's mass should be large compared to the
3-brane tension. We leave this criterion open since it is highly
model-dependent.
\end{itemize}

\begin{figure}[h]
\begin{center}
\includegraphics[width=8cm]{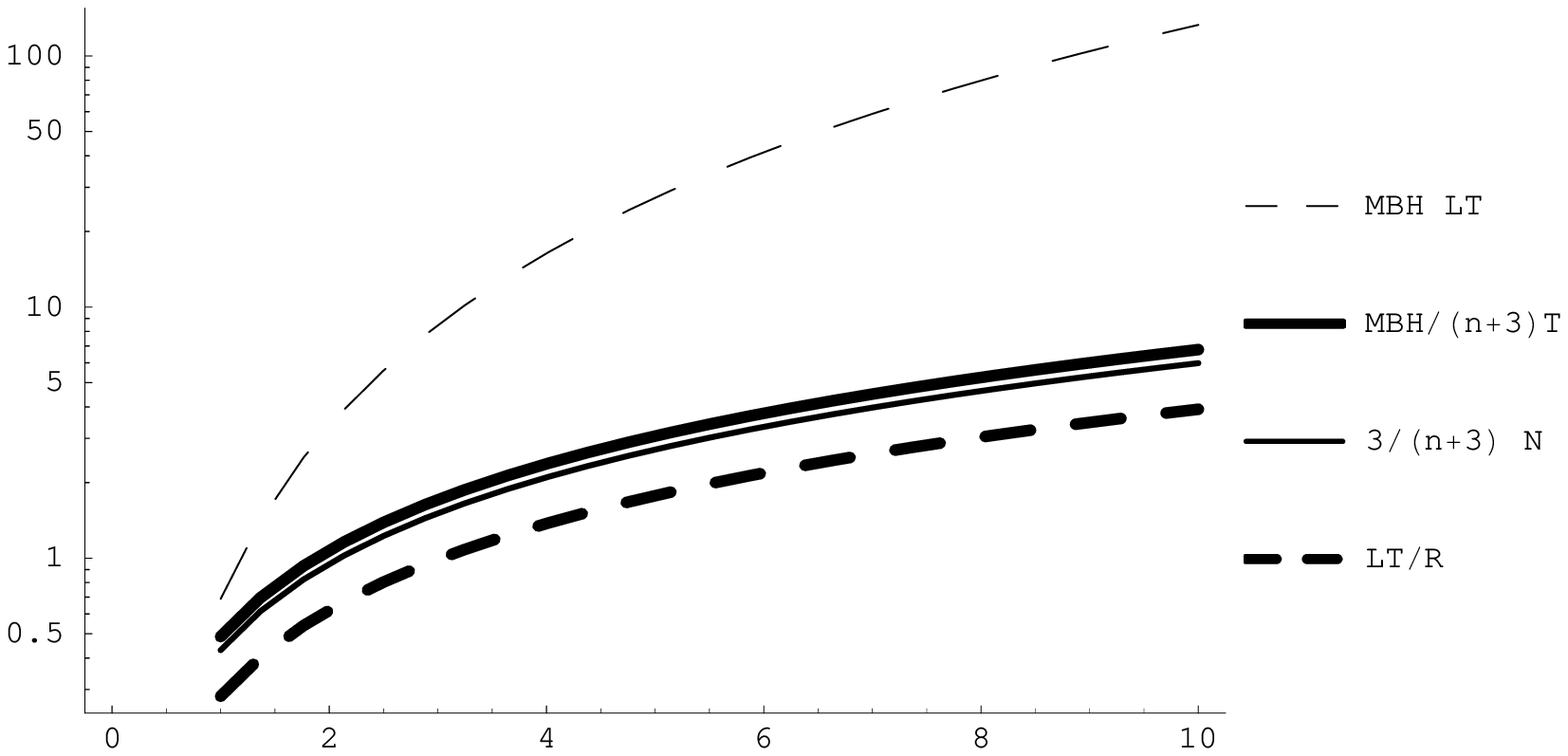}\includegraphics[width=8cm]{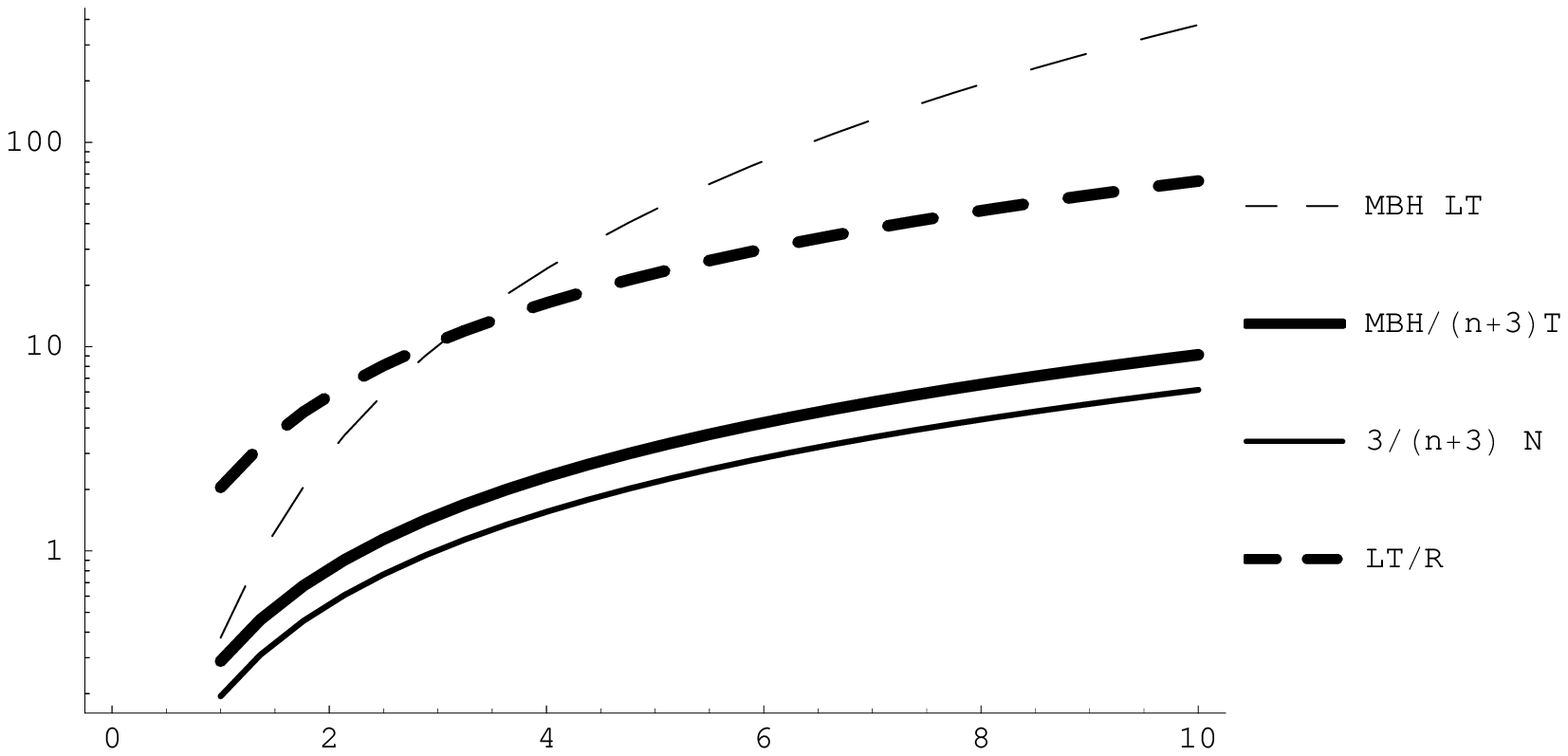}
\end{center}
\caption{Possible criteria for $x_{min}$ plotted as a function of
$x_{min}$.  ADD with n=6 is plotted on the left and RS is plotted
on the right.} \label{critplot}
\end{figure}

The strongest criteria are plotted in Figure~\ref{critplot} as a
function of $x_{min}$ (with the exception of Schwarschild vs.
Compton wavelength which would just be a vertical line) where the
ratios are chosen such that every curve plotted should be greater
than one if the criteria is satisfied.  These criteria highlight
the uncertainty in defining a precise threshold, and also indicate
the blackhole threshold might be well above the putative Planck
scale. We stress here that even though the various criteria might
be satisfied for $x_{min}\gtrsim 3$ or 4 (except for the
wavelength criterion), all these criteria are should really be
held to being $\gg 1$ not just $\sim 1$ in which case $x_{min}$
should be much larger in principle.  They also show that the
values of $x_{min}$ that were used in previous analyses
\cite{Giddings:2001bu,fenginelast} might be  too low to trust to
be in the thermal regime (and of course brings into question those
analyses that neglected $x_{min}$ entirely). As we will see
however, higher values of $x_{min}$ yield too low a production
rate to appear at the LHC.

\subsection{Inelasticity}
  In addition to the thermality criteria above that raise the black hole
energy threshold, another critical effect is energy loss of the
colliding partons before their energy is trapped behind a black
hole horizon.
  One of the most important effects is to understand exactly how much
energy of the initial parton parton system ends up going into the
mass of the intermediate black hole.   We can define an
inelasticity parameter as in~\cite{fenginelast} $y\equiv
M_{BH}/\sqrt{\hat{s}}$ which when less than 1 requires probing the
PDFs at larger x and thus reducing the cross section possibly by
many orders of magnitude compared to initial
estimates\footnote{There are other effects that modify the cross
section, i.e. the maximum impact parameter that can still create a
black hole in comparison to the Schwarzschild radius and
$\mathcal{O}(1)$ factors in front of the putative cross section
$\sigma\approx \pi r_S^2$ however for the LHC these effects are
not nearly as crucial as the actual mass scale that defines the
black hole production.}.

\begin{figure}
\begin{center}
\includegraphics[width=8cm]{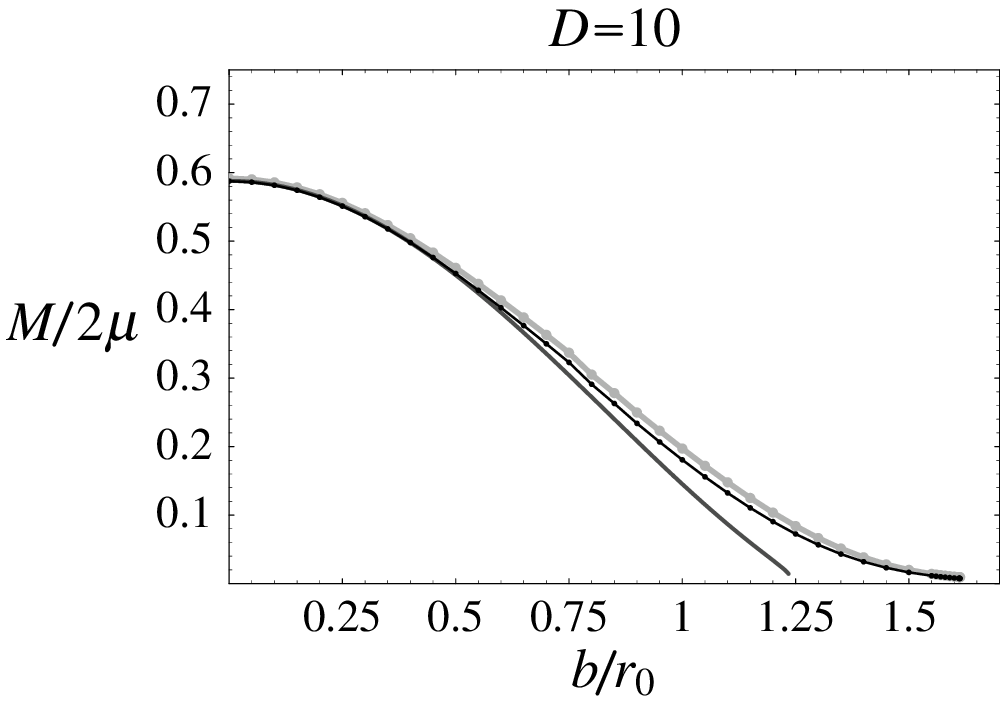}\includegraphics[width=8cm]{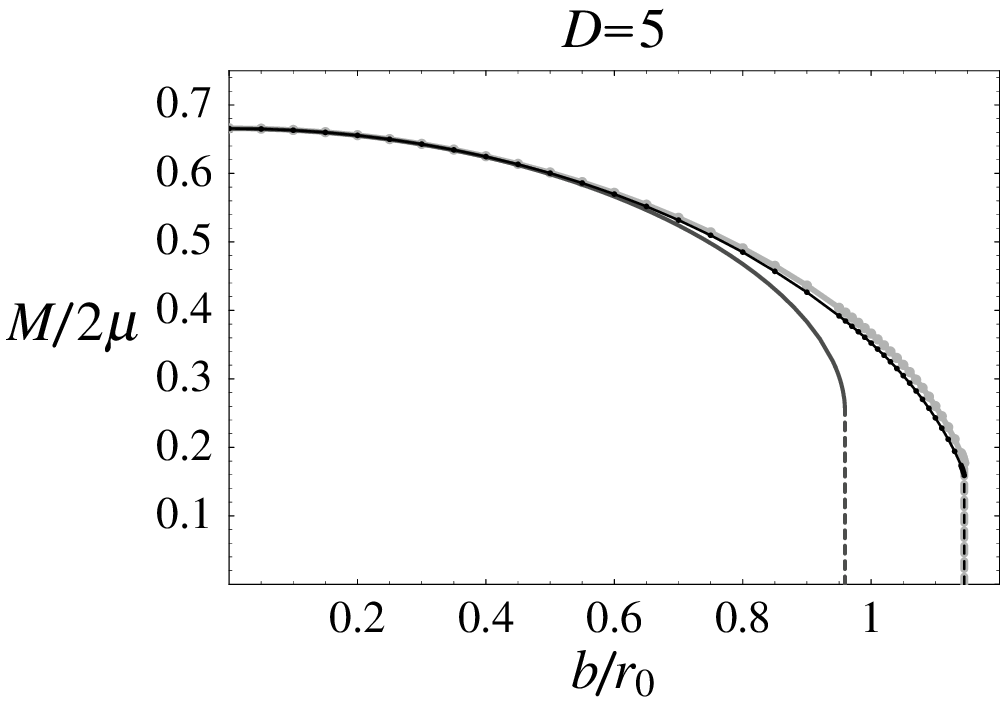}
\end{center}
\caption{From Fig.10 of \cite{Yoshino:2005hi}.  The ratio of the
mass of the putative black hole compared to the initial energy of
the collision is plotted as a function of the impact parameter
divided by a unit $r_0$ that approximates the Schwarzschild radius
if all the energy of the initial collision were to end up as a
black hole.  The lowest curve represents the calculation
of~\cite{Yoshino:2005hi}, and previous estimates
from~\cite{Eardley:2002re,Yoshino:2002tx} are also included.}
\label{yoshrych}
\end{figure}

The program of calculating this inelasticity goes back to
unpublished work of Penrose and the work of D'eath and
Payne~\cite{D'Eath:1992hb,D'Eath:1992hd}, who examined in four
dimensions and zero impact parameter, the fraction of energy
emitted in gravitational waves when colliding to Aichelburg-Sexl
shock waves representing two highly boosted massless particles.
This work was extended to extra dimensions and non-zero impact
parameters by the seminal work of Eardley and
Giddings~\cite{Eardley:2002re} and then further refined
by~\cite{Yoshino:2002tx,Yoshino:2005hi}.  In Figure~\ref{yoshrych}
we present the relevant results of~\cite{Yoshino:2005hi}, for the
ratio of the mass trapped in the apparent horizon compared to
initial energy as a function of the impact parameter for a 10
dimensional black hole (hereafter referred to as ADD) and 5
dimensional black hole (hereafter referred to as RS) which are
relevant for our discussion.

As one can see from Figure~\ref{yoshrych} the largest energy
fraction entering the black hole for both ADD and RS is
$\mathcal{O}(.6)$ occurring for zero impact parameter. However
they have different functional dependencies with respect to the
impact parameter, and the ADD fraction goes down to $y\approx 0$
while RS goes to about $y \approx .2$ at the largest possible
impact parameters where an apparent horizon still forms. These
estimates are interpreted as lower bounds on the inelasticity but
we stress that they are also calculated classically and for
energies that are approaching the Planck scale it is not obvious
how this will be modified.

To quantitatively include this inelasticity, we need to include
the impact parameter dependent effect of inelasticity in
calculating the black hole production cross section.  Implicitly
when calculating the cross section of a proton proton event we
have summed over the possible impact parameter already when using
the parton parton cross section
\begin{equation}
\sigma (pp\rightarrow X)=\sum_{i,j}\int dx_1 dx_2 f_i (x_1) f_j
(x_2) \sigma (ij\rightarrow X).
\end{equation}
To include the effects of inelasticity we adopt the impact
parameter weighted average of the inelasticity used in
~\cite{fenginelast}
\begin{equation}
\sigma (pp\rightarrow BH) \equiv \sum_{i,j} \int_0^1 2z dz
\int_{\frac{(x_{min} M_D)^2}{y(z)^2 s}}^1 du \int_u^1 \frac{dv}{v}
f_i (v,Q) f_j (u/v,Q) \sigma_{i,j\rightarrow BH} (M_{BH}= us),
\end{equation}
with $z=b/b_{max}$.  The function $y(z)$ is given in our case by
the results of~\cite{Yoshino:2005hi}, as shown in
Figure~\ref{yoshrych}.  This weighting of the impact parameter
obviously shows a difference between the RS and ADD cases, because
in 10 dimensions the inelasticity parameter is smaller at order
one impact parameters, meaning relatively higher energy will be
needed to make a black hole.
\begin{figure}
\begin{center}
\includegraphics[width=8cm]{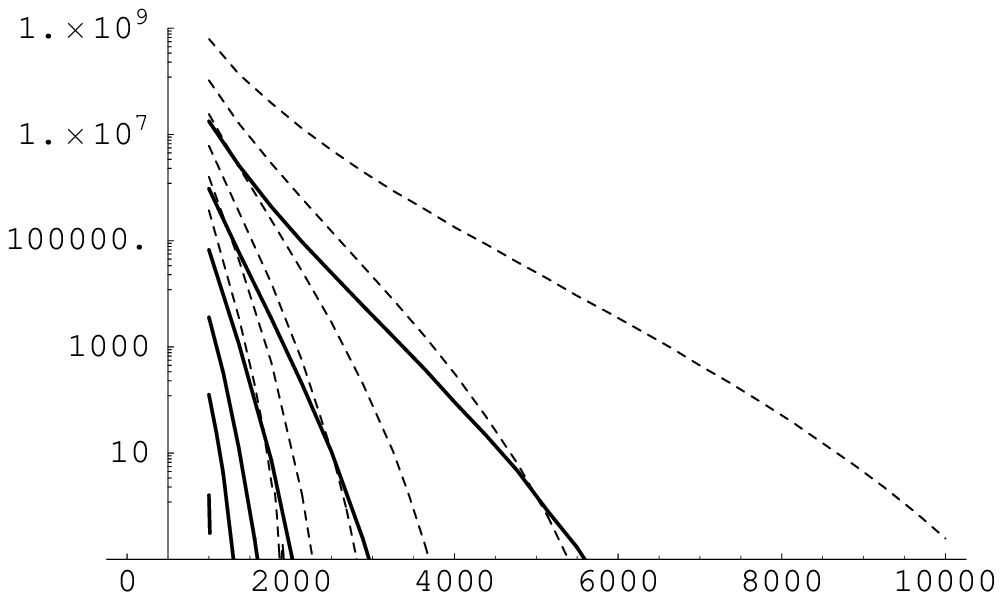}\includegraphics[width=8cm]{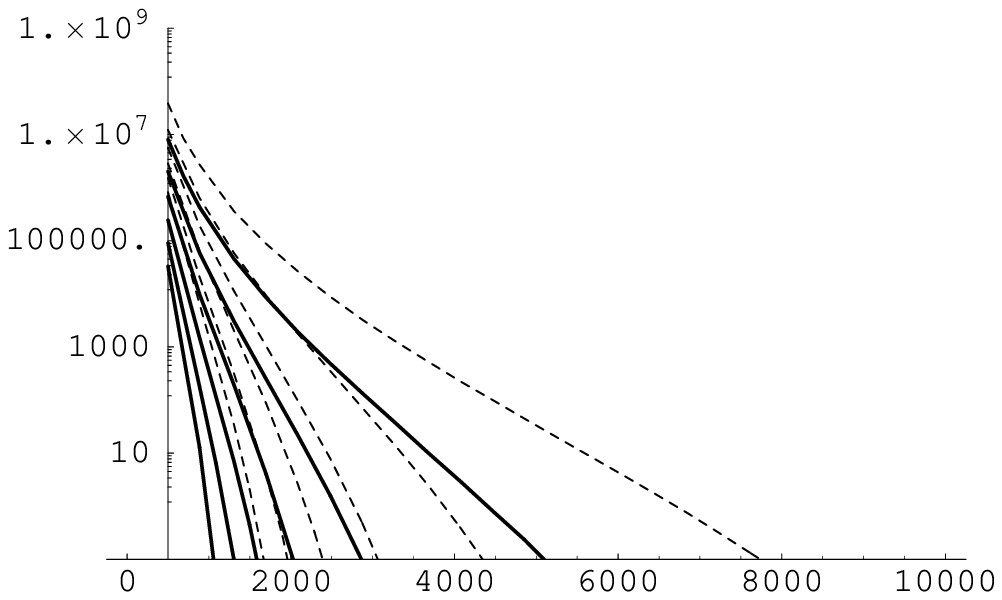}
\end{center}
\caption{Total black hole cross section in femtobarns,
including(solid curves) and not including(dashed) inelasticity as
a function of $M_D$ for ADD with $n=6$ and $\tilde{M}$ for RS1.
The different curves from highest to lowest correspond to
$x_{min}=1-6$.} \label{addrstot}
\end{figure}
The total black hole cross section with and without inelasticity
for both ADD and RS is shown in Figure~\ref{addrstot}.  As
demonstrated in Figure~\ref{addrstot} the inclusion of
inelasticity can reduce the total cross section by several orders
of magnitude, which is consistent with the results
of~\cite{fenginelast} who used~\cite{Yoshino:2002tx} to define
their inelasticity.  It is interesting to note that these effects
are more important for ADD than RS in terms of reduction of total
cross section, as it is interesting that the inelasticity is higher
for lower dimensions. While the rates presented in
Figure~\ref{addrstot} for the inclusion of inelasticity are taken
as a lower bound for the black hole cross section, one should keep
in mind that $x_{min}$ lower than the criteria presented in
Section~\ref{criteria} have been plotted and it is unclear what
the ``effective" inelasticity will be when quantum gravity effects
are taken into account.

\section{Black Hole Decays}

  In the previous sections we have argued that it is unlikely that the LHC
will produce thermal black holes, since the thermality criteria
require a black hole threshold above the putative
higher-dimensional Planck scale and furthermore energy is lost
through initial radiation. In this section we go a step further
and argue that even if black holes were produced, they are rarely if ever in
a regime where they will produce the ``fireball" explosions
consisting of a high multiplicity isotropic distribution of
particles that are the most highly emphasized
\cite{Dimopoulos:2001hw,Giddings:2001bu} black hole signature and
possibly  even revealing the negative specific heat that characterizes black holes.

Since this signature relies on high multiplicity events, it is
worth checking over what parameter range one expects to find high
multiplicities. Although not necessarily reliable for low
multiplicities, we quantify this consideration by exploring the
average number of particles assuming standard classical black
holes with a thermal distribution of final state particles obeying
Poisson statistics\cite{fenginelast,Bekenstein:1995ju} to
determine the fluctuation about this mean value. The point is to
show the relative merits of low and high multiplicity states. We
use as a target ``high multiplicity" six or more particles.
Although far from a fireball, we are trying to allow the most
optimistic assumption for a multiparticle state. We compare this
reach to two body final states in the figures below.

In Figure~\ref{addrs6} we plot the cross sections with and without
inelasticity for both 6 or more particles(multiparticle) and 2
particles.  To summarize and better demonstrate the relative
potential strengths of multiparticle vs. two particle final states
we plot in Figure~\ref{add6v2} the region in parameter space for
the multiparticle and 2 particle final states with a .1 fb cross
section.

\begin{figure}[h]
\begin{center}
\includegraphics[width=8cm]{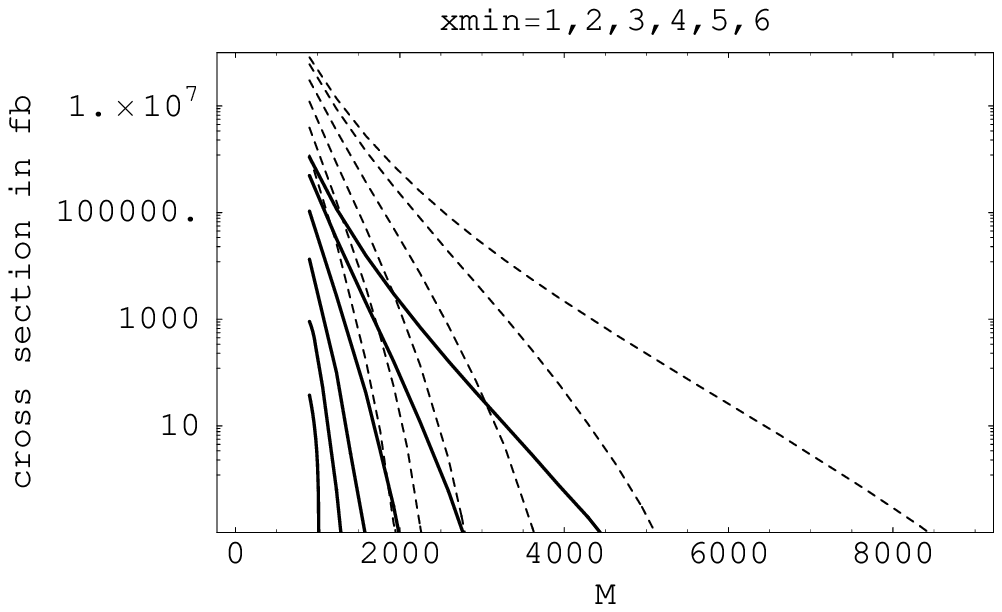}\includegraphics[width=8cm]{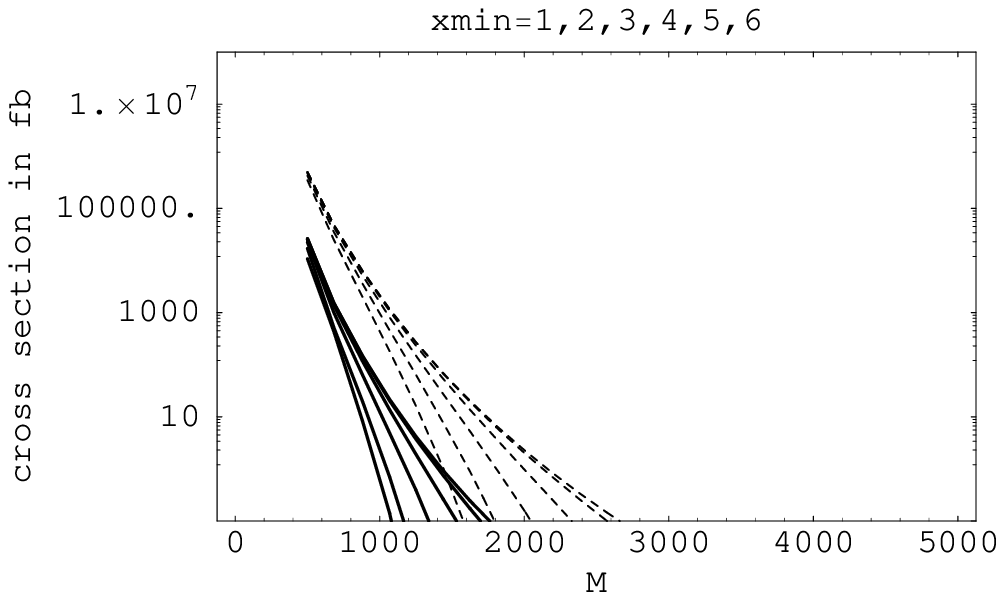}\\
\includegraphics[width=8cm]{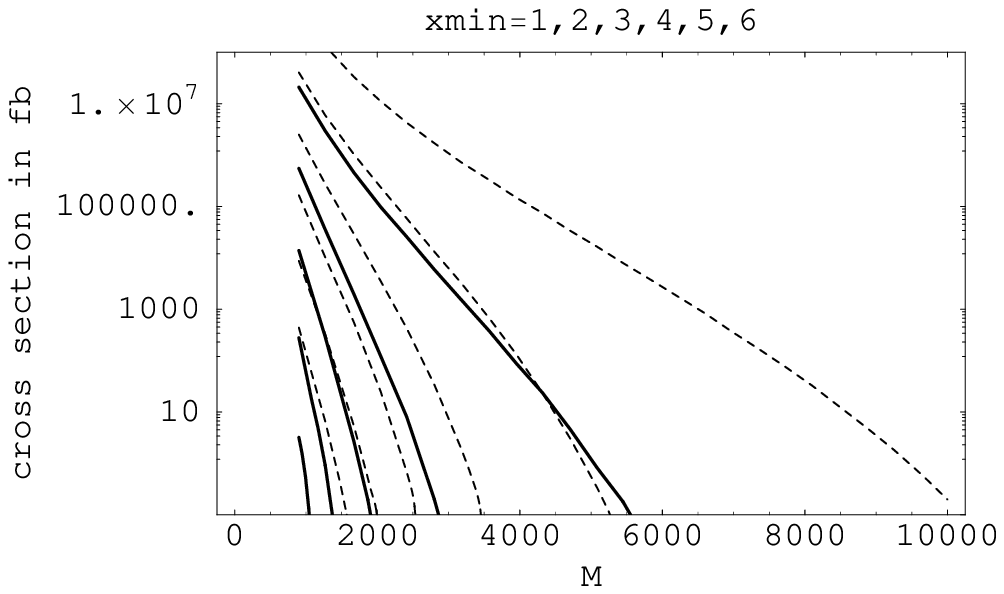}\includegraphics[width=8cm]{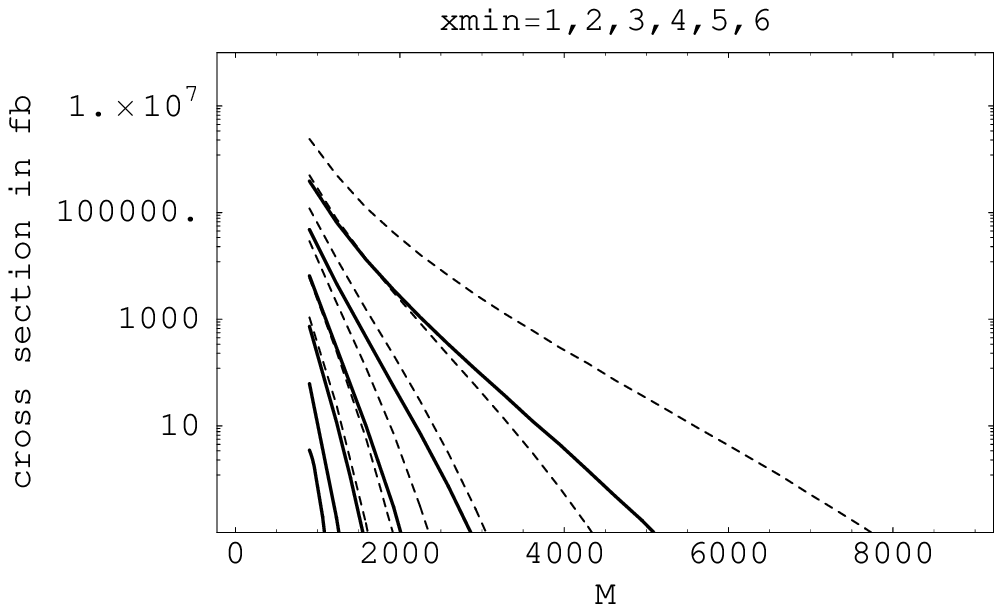}
\end{center}
\caption{In the upper plots curves of total cross section for
having 6 or more particles, including(solid curves) and not
including(dashed) inelasticity as a function of $M_D$ for ADD with
$n=6$ and $\tilde{M}$ for RS1.  The different curves from highest
to lowest correspond to $x_{min}=1-6$.  In the lower plots the
same curves are plotted for having 2 particles instead of 6 or
more.} \label{addrs6}
\end{figure}

\begin{figure}[h]
\begin{center}
\includegraphics[width=8cm]{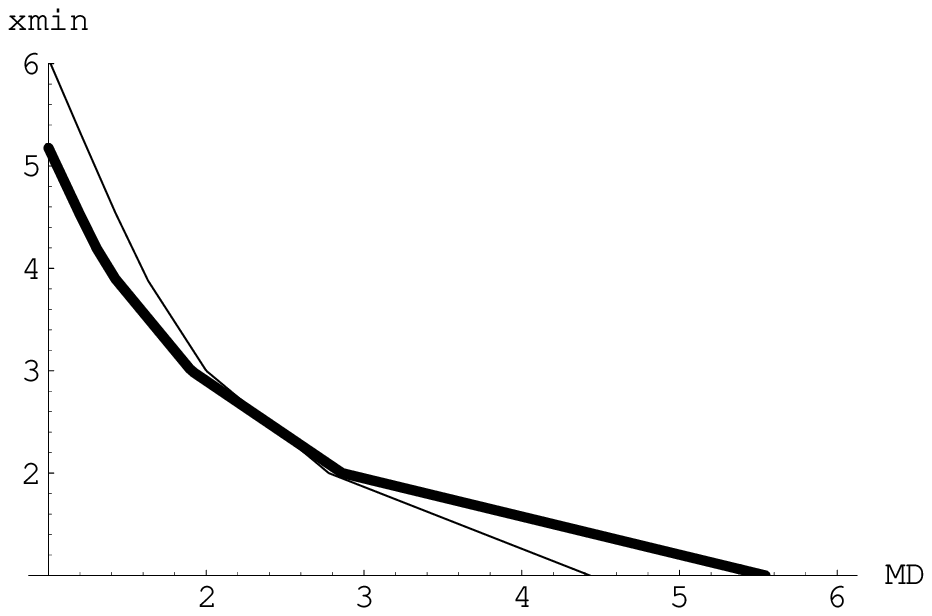}\includegraphics[width=8cm]{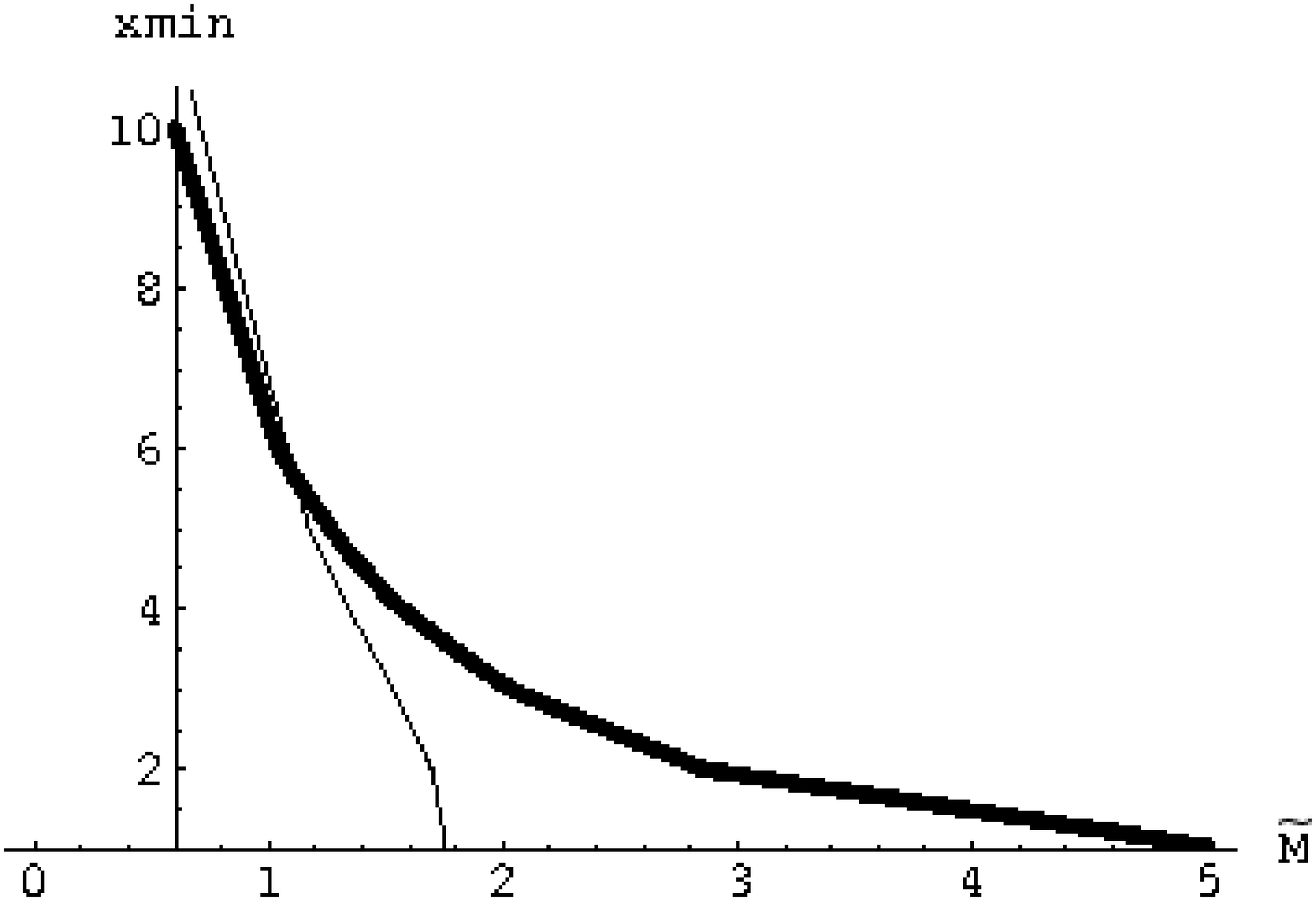}
\end{center}
\caption{Curves of constant .1 femtobarn cross section including
the effects of inelasticity and a probability for getting either 2
particles(thicker curve) or greater than 6 particles(thin curve).
In the left hand panel the curves are for ADD with 6 extra
dimensions and are plotted as a function of $x_{min}\equiv
M_{BH}/M_D$ and $M_D$.  In the right hand panel the curves are for
RS1 as a function of $x_{min}\equiv M_{BH}/\tilde{M}$ and
$\tilde{M}$.} \label{add6v2}
\end{figure}

We see that the ``reach"\footnote{Here we are defining reach as
just a possible observation of signal assuming the background is
non-existent.  This is not meant to be a statistically significant
reach; nevertheless it gives one the potential reach if
backgrounds are under control.  In the next section for the two
particle states we will show backgrounds in the 2 particle state.}
of two particle final states is in all cases at least as good as
the multiparticle final state. Therefore a study of low
multiplicity final states  might explore black hole-like objects
even when $x_{min}$  is not high enough to guarantee a thermal
final state or a black hole.

To be as optimistic as possible, we also
checked for the maximum number of particles assuming a .1 fb cross
section according to a Poisson distribution for a given
$\tilde{M}$$[M_D]$ where we looked for the maximum number of
particles with that cross section.  The maximum particle number in
the RS case with a .1 fb cross section was on the order of 20 for
$\tilde{M}=500$ GeV and is about 9 for $\tilde{M}=1$ TeV , and is
only 6 for $\tilde{M}$ of 1.5 TeV.  The maximum particle number in
the ADD case for $M_D=900$ GeV  was about 20, for 1.4 TeV was
about 14, and for 1.9 TeV was about 10. Although the later case
might sound adequate, it should be kept in mind that this number
depends on decays onto the brane. If we asked about the
distribution of energy among thermal bulk particles, that is how
many bulk particles would we expect for this sized black hole, the
answer would be divided by 3. And this was for the best possible
cases. So the black hole signature is not likely to be an
isotropic burst of a large number of particles. Instead we expect
low multiplicity final states to dominate.

Given the relative weakness of the
muliparticle final states the likely black hole signature will not
be an isotropic burst of a large number of particles. Instead we
expect low multiplicity final states to dominate. We consider the
consequences of this conclusion in the next section.

\section{Two Body Final States} \setcounter{equation}{0}
\setcounter{footnote}{0}

Examining the formulae for the average number of particles emitted
in the decays~(\ref{decay1}) and (\ref{decay2}), we see that for
RS only for $M_{BH}>4\tilde{M}$ is the average number of particles
emitted greater than 2, and for ADD you need $M_{BH}>1.5 M_D$.
Clearly for $x_{min}$ satisfying the criteria we've listed this is
not a problem. But it makes manifest that for low $x_{min}$ our
``black holes"  decay into only a small number of particles.
However, even if the decay is not a true thermal black hole, some
interesting new signature is likely to appear and could be a
valuable indicator of strong gravity effects-one whose reach in
almost all cases is comparable to or exceeding the reach of the
multiparticle final states.

   We now consider the implications of black holes, or
other quantum gravity effects, for 2$\rightarrow$2 scattering
processes. Whether or not true black holes appear at center of
mass energies of order the Planck mass, we expect that true or
virtual black holes or simply strong gravity effects will lead to
an increase in the 2$\rightarrow$2 production cross section as we
approach the Planck scale, as can be seen from the large cross
sections in the previous section.  Later on when we include
backgrounds this increase will become even more manifest. In
practice, because the calculation is inherently nonperturbative in
this regime, we cannot precisely calculate the scattering.
However, by considering a few examples we show  that under
reasonable assumptions we can gain insight into quantum gravity by
studying these processes.

In principle exception to enhanced two-body production might be weakly coupled string
theory. But in this case we would see the string states or other
effects (see below) well before the black hole scale, which in any
case would be out of reach (since it is of order
$M_S/g_s^2$~\cite{bhstring}).

  So rather than explore only the tail of black hole distributions where
multiparticle states could dominate, we  examine the Planckian
``black holes"  (by which we mean any quantum gravity effect or resonance) where $M_{BH}\sim M$.    Given the PDF fall offs at
the LHC and the flux limitation of UHECR experiments black holes
will be dominantly produced at the lowest mass available to them
in any foreseeable experimental setting that they could be
potentially produced in. Taking into account inelasticity amongst
other effects  we don't expect high mass black holes, and one
really needs to explore as low of mass ``black holes" as possible
within the context of the most quantitative statements that can be
made.

Furthermore, we have seen by considering the maximum particle
number that even under the most optimistic assumptions on scales
and threshold, we are unlikely to create truly thermal high
multiplicity black holes but instead low multiplicity states.

The most dramatic two body final state signature one might hope to
find would be one that violates global quantum numbers, such as
$\mu$,$e$. However, since this physics occurs at the TeV scale,
there are already strong constraints since turning around such a
process would presumably permit flavor-changing lepton decays for
example. This might occur through virtual black hole exchange or
directly through the dangerous TeV-scale physics. The latter could
be suppressed if there are effectively large anomalous
dimensions---that is the operator turns on only at high energies.
The former could be suppressed since we don't understand virtual
black holes. But we conservatively assume that there is either
separation of particles in the higher dimensions or a
spontaneously broken gauge symmetry so that such dramatic final
states will not occur. Of course if they are seen, we would have
to rethink the loopholes to check whether such events could arise
from black holes.

However, even if black holes don't provide dramatic global quantum
number-violating decays,  we expect an observable signal. We focus
first on the two jet signal, but we will also show that the two
lepton final state can be very helpful in distinguishing among
quantum gravity models. One reason we first focus on the jet final
state is that we don't know how gauge charge is shed. Since the
dominant parton initial state will be quark-quark or gluon-gluon
or gluon-quark, we need to know how gauge charge flows. If it is
shed in soft quark or gluons from the initial partons, a neutral
black hole could be produced in which case the two jet final state
would be expected at about ten times the rate for leptons, though
the lepton background is smaller so more detailed studies of
leptons can be performed even with lower cross section.  However,
if the initial state carries gauge charge (remember we are dealing
with low entropy black holes that decay instantaneously), we would
expect a two jet but not a lepton  final state.

However, even though the cross section might increase over that of
the Standard Model, it might appear that finding black holes or
even effects of strong gravity will be difficult if they are only
revealed in two body final states since the two jet background
would have to be rather precisely predicted. We show that because
the jets will have a much more transverse distribution than QCD
background, which is dominated by t-channel exchange, the new
events would be readily distinguishable. In fact, this difference
in angular distribution is a feature of any contact operator due
to strong interactions. Because we are not yet safely in the
classical gravity regime, we will also consider the possible
interpretation of black hole and strong gravity effects in terms
of higher-dimensional contact terms generated by strong dynamics.
We now exploit this similarity to suggest a new way to search for
interesting effects from quantum gravity. This 2 body final state
is interesting since there is almost certainly a bigger reach than
for multibody final states which are in any case very unlikely to
be thermal and because this is truly the quantum gravity regime
where classical predictions don't apply. Although we can't predict
the results from first principles, the measurements at the LHC can
in principle distinguish different models of quantum gravity as we
demonstrate below.

With QCD, the 2$\rightarrow$2 scattering cross section at high
energies is very forward peaked because of $t$-channel gluon
exchange. When looking for new physics, it is therefore useful to
look at both $d\sigma/dM$ and the angular distribution of the
decay products. To optimize the search for deviations from the
standard model, it is useful to define a quantity
$\reta$~\cite{Abbott:1998wh}, which is the ratio of the number of
events with pseudorapidity between 0 and 0.5 divided by that for
pseudorapidity between 0.5 and 1.0. $\reta\equiv
N_{events}(0<\vert\eta\vert)<.5)/N_{events}(.5<\vert\eta\vert)<1)$
  Deviations from the
asymptotic QCD value of 0.6 would indicate new physics.  The
quantity $\reta$ is useful because in measuring $d\sigma/dM$ there
are a great deal of systematic uncertainties coming from for
instance understanding the jet energy, resolution etc, which means
that in searching for new physics such as compositeness in dijets
$d\sigma/dM$ is not necessarily a reliable quantity.  However in
the ratio $\reta$ most systematic effects cancel and thus the
error is reduced to being essentially statistical only.  The
variable $\reta$ which originally was used at
D0~\cite{Abbott:1998wh} has thus been carried over for LHC studies
at CMS~\cite{cmsdijet}.

We will use this description to interpolate between a notion of
some new strongly coupled physics  prior to true thermal black
hole formation compared to just a sharp turn on of classical black
hole production. Clearly the two body final state cross section is
enhanced by strong gravitational effects, even before we reach the
true black hole threshold.

In fact, virtual black holes are only one type of quantum gravity
effect that might lead to changes in the 2$\rightarrow$2
scattering cross section. We now list some possibilities, consider
constraints in the following section, and how experiment might
distinguish among the possibilities in the sections that follow.

We will also consider the role that lepton final states can play
in distinguishing among possibilities.

\subsection{Quantum Gravity and 2$\rightarrow$2 Scattering}

No matter what the theory of quantum gravity, the 2$\rightarrow$2
scattering cross section might well be the first clue of low-scale
quantum gravity and can furthermore yield insight into quantum
gravity behavior. Our point is not that any one of the behaviors
we consider necessarily applies but that we should be able to
experimentally distinguish among them according to the
differential cross section and angular distribution of both jet
and leptonic final states. There is a great deal of physics that
can be done with dijet final states that has been largely
neglected up to now. We now consider several possibilities.

  Even knowing nothing a priori about
quantum gravity   it is difficult to imagine no enhancement of the
two body final state cross section at scales close to those at
which the black hole cross section turns on.
   If a string theory description does not apply, one
might expect a particle description does. The only way to avoid
two body final states would be an instantaneous decay into some
minimum number of particles. But if we describe the decay through
a higher dimension operator as might be appropriate in a particle
description, it is clear that the operator coefficient would be so
enhanced that even closing off the final states through loops to
make a two particle final state, there would still be a sizable
decay into two body final states. If a weak string description
applies, we expect effects of the sort we soon consider. If,
however, string theory is strongly coupled, we cannot predict the
behavior but can again see what experiment might tell us. In this
case it is reasonable to expect that 2$\rightarrow$2 processes are
enhanced as we approach the black hole scale, where we mean the
scale at which strongly interacting gravity gives rise to truly
thermal black holes. At smaller energies it is reasonable to
expect hard scatterings due to multigraviton exchange.

Our discussion of two body final states as an indication of black
hole production contrasts with Ref. \cite{banksfischler}, which
argues that the 2-body final state is diminished when high
multiplicity final states dominate black hole decay. We do not
dispute this conclusion applies at high energy, but point out we
expect a range of energy at or about the Planck scale for which
the two body final state dominates and increases the
2$\rightarrow$2 cross section over that of the Standard Model.

\subsubsection{Dijet ``Black Holes"}\label{sec:dibh}

According to black hole formulae, we expect an increase in the
2$\rightarrow$2 cross section near the black hole threshold. Of
course, eventually multiparticle states would dominate as black
holes are produced at sufficiently high energy, cutting off the 2
body final state \cite{banksfischler}. But sufficiently close to
threshold we don't expect this to happen. Even though the
classical formula does not apply and we don't truly expect a black
hole, we expect enhancement of 2$\rightarrow$2 that we model
according to the black hole cross section near threshold. This
gives

\begin{equation}\label{2bodeq}
\sigma(\sqrt{\hat{s}}>x_{min} M)\approx \pi r_S^2 P_2
\end{equation}
\begin{equation}
P_2=e^{-\langle N\rangle} \sum_{i=0}^{2} \frac{\langle
N\rangle^i}{i!}
\end{equation}

Note that other authors \cite{Giddings:2001bu} treat the final
decay as a 4 stage process, with balding, spindown, Hawking
radiation, and the final explosion.  In practice however the
existing black hole generators~\cite{catfish,charybdis,truenoir}
only incorporate hawking radiation and the final evaporation. Our
point of view is that ``black holes" produced near threshold decay
instantaneously, where the average number of particles can be
approximated by a Poisson distribution
\cite{Bekenstein:1995ju,fenginelast}. Note that the most recent
black hole generators~\cite{catfish,charybdis} always have at
least two particles in the final state of the decay (unless a
remnant is postulated as an option as in~\cite{catfish} a
possibility we think unlikely given the results
of~\cite{remnants}), so they will never have the two body final
states we are looking at unless Hawking evaporation is entirely
absent (\cite{truenoir} leaves this as a parameter and can
potentially examine 2 body final states). The probability for
$\langle N \rangle$ we assign, which is computed assuming Hawking
radiation and a Poisson distribution, is probably not accurate.
Nevertheless it is a rough approximation to the real probability
of a $2\rightarrow 2$ process that we expect to occur and in some
sense a conservative estimate since we do not demand that the
probability is $1$ for low energies.

We note that we don't know how to treat interference since we are
in a nonperturbative regime. In our results below, we simply add
the Standard Model and black hole cross sections.

\begin{figure}[h]
\begin{center}
\includegraphics[width=8cm]{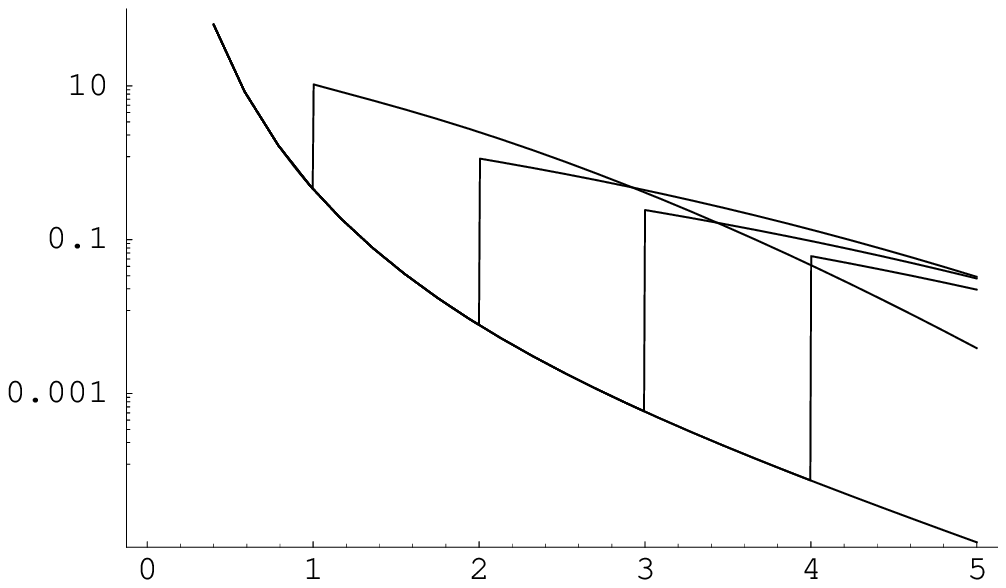}\includegraphics[width=8cm]{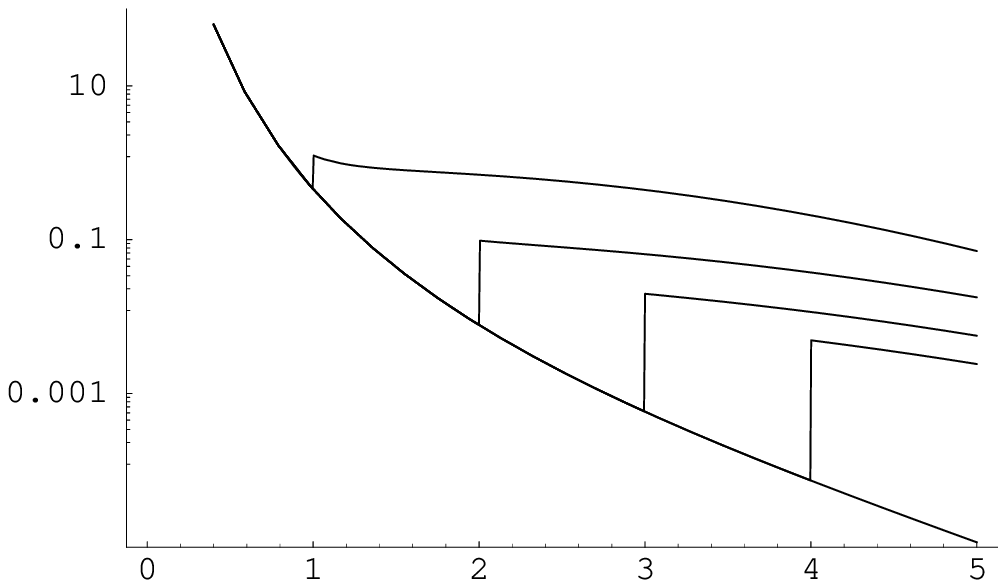}\\
\includegraphics[width=8cm]{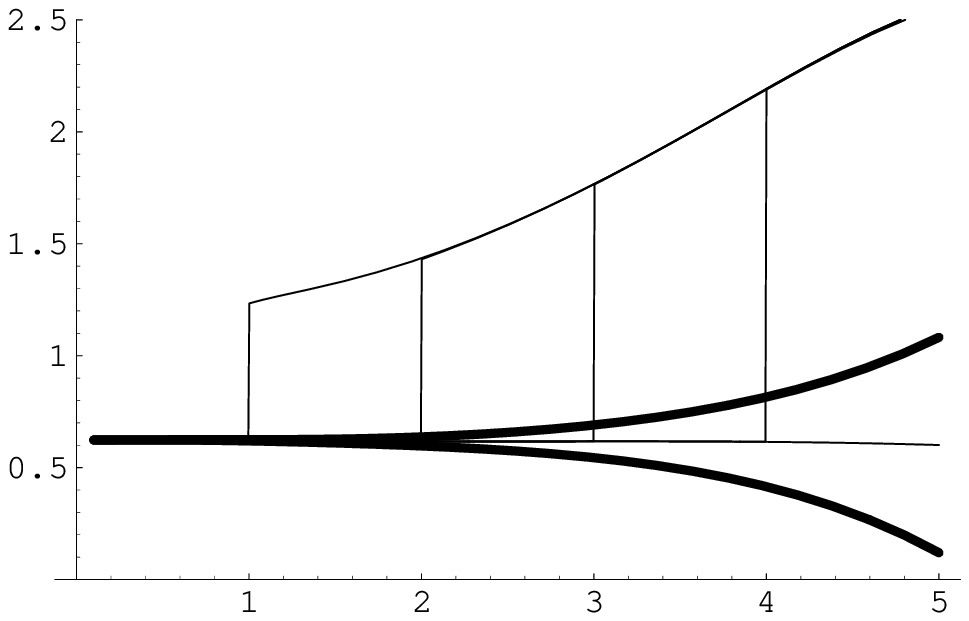}\includegraphics[width=8cm]{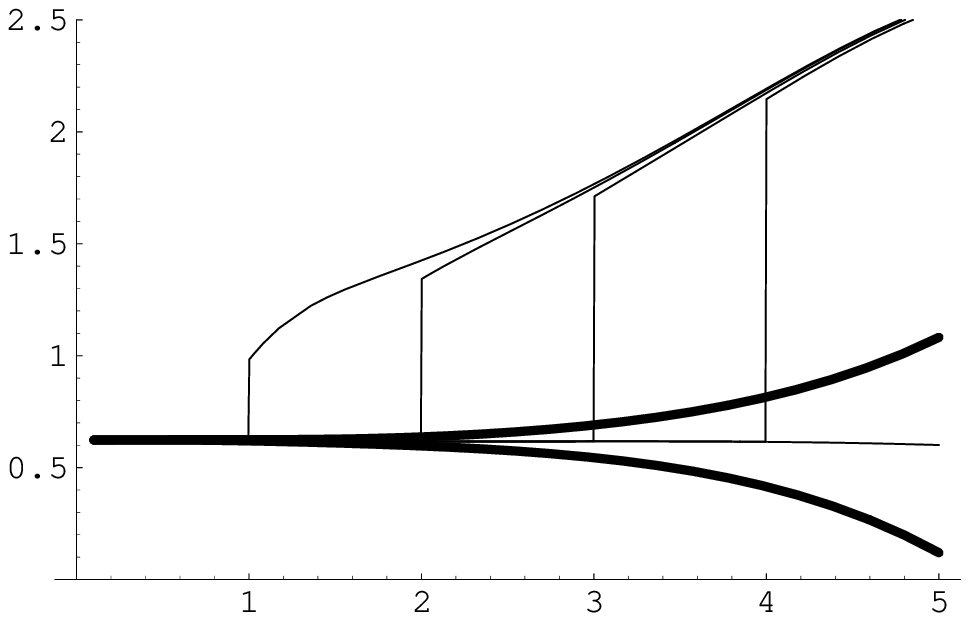}
\end{center}
\caption{In the upper plots $d\sigma/dM_{jj}$(units of pb/GeV) vs
$M_{jj} (TeV)$ is plotted for the case of SM QCD background, and a
n=6 ADD model ``black hole" behavior with $M_D$=1,2,3,4 TeV and
$x_{min}=1$ in the lefthand plot and a RS1 black hole behavior
with $\tilde{M}=1,2,3,4$ TeV and $x_{min}=1$ in the righthand
plot.  For other values of $x_{min}$ the curves simply start at
the corresponding dijet mass.  In the lower two plots the $\reta$
is plotted for the same parameters.} \label{twobodybh}
\end{figure}

Note the distinctive features of this model visible in
Figure~\ref{twobodybh}. First of all, we see that the cross
section turns on suddenly. Of course this is a consequence of our
approximation of sudden turn on at threshold but even allowing for
some smoothing we would expect a more dramatic rise in  cross
section than with higher dimension operators for example as we
will show below.  The rapid rise would be expected if the black
hole is a resonance, or a convolution of a continuum of resonances
that might occur near the quantum gravity scale.

   This rapid rise in cross section is  mimicked in
the parameter $\reta$ which measures the angular distribution. We
would see the QCD value of 0.6 suddenly jump to a larger value,
indicative of a much more transverse distribution. As expected,
the two jets due to black holes are far more transverse than the
QCD background. The more rapid deviation from QCD could help
distinguish black-hole type behavior from other strongly
interacting physics. Of course, this rapid turn on was based on
our assumption that the black hole event rate takes over at
$x_{min}=1$ (here we mean just our 2 body final state and not the
true thermal black hole). In reality, we expect a smoother
interpolating behavior. Nonetheless, it would be very bizarre
strong physics other than gravitational that would have a sudden
(or even smoothed out) jump at higher energies. One would need a
model of strongly interacting physics that turns on in the UV but
whose effects disappear in the IR.

\begin{figure}[h]
\begin{center}
\includegraphics[width=8cm]{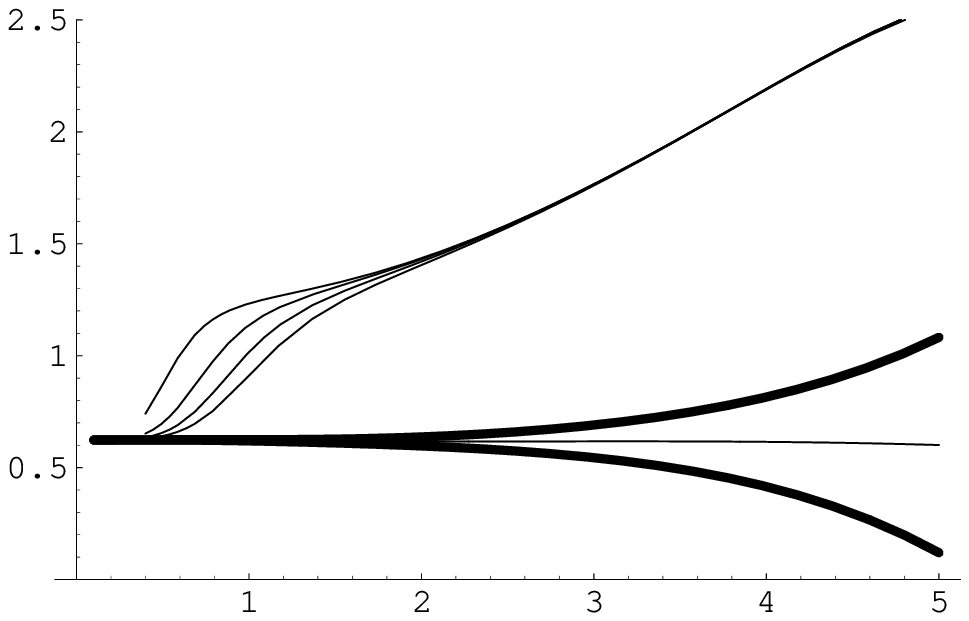}\includegraphics[width=8cm]{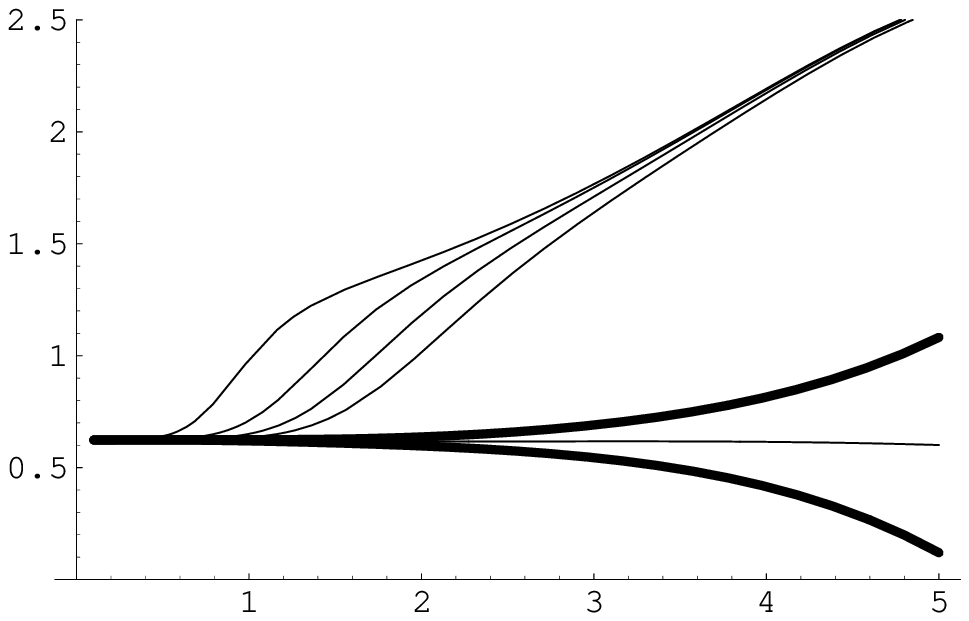}
\end{center}
\caption{In the upper plots $d\sigma/dM_{jj}$(units of pb/GeV) vs
$M_{jj} (TeV)$ is plotted for the case of SM QCD background, and a
n=6 ADD model ``black hole" behavior with $M_D$=1,2 TeV and
$x_{min}=1$ in the lefthand plot and a RS1 black hole behavior
with $\tilde{M}=1,2$ TeV and $x_{min}=1$ in the righthand plot.
For other values of $x_{min}$ the curves simply start at the
corresponding dijet mass.  In the lower two plots the $\reta$ is
plotted for the same parameters.} \label{twobodynocut}
\end{figure}

The apparently discontinuous jump in $\reta$ reflects the fact
that when the ``black hole" cross section turns on it dominates
QCD and there is no interference.  If one were to extrapolate
below $x_{min}=1$ to the regime where interference effects could
potentially be visible one finds a different behavior as seen in
Figure~\ref{twobodynocut}.  In Figure~\ref{twobodynocut} with no
$x_{min}$ value chosen, one sees that at high energies various
values of $\tilde{M}[M_D]$ look identical whereas there is a
difference at low dijet masses.  This is because even including SM
processes black holes dominate at high energies and the angular
dependence is that of an isotropic distribution governed by the
PDFs.  At low dijet masses if there were interference with the
black hole cross section~(\ref{2bodeq}) QCD is competitive and one
can see the different scaling of $E/\tilde{M}[M_D]$ for different
choices of the Planck scale.

\subsubsection{Stringy Behavior}

  We have so far assumed the string coupling is of order unity
and that black hole like behavior will appear without any obvious
signs of a string theory regime.  However, if the string scale is
low and the coupling is weak, string theory could give rise to
resonances that would change the differential cross section and
$\reta$.

  In the case of weakly coupled string theory it is well known that above
the string scale, the two body final state is reduced
exponentially at any large transverse angle. In practice, we
expect power law suppression for values of $g$ that are not too
small since for sufficiently high genus the loop string
contribution will no longer be exponentially
suppressed~\cite{grossmende} (the behavior is of the form $g^{2G}
s^{G+1} e^{-s/(G+1) f(c)}$, where $f(c)$ gives the dependence on
angle).  Nonetheless, we do expect a dip at high energies that we
will explore.

Given a string realization of higher dimensional gravity, be it
ADD or RS, there could be additionally a regime of string ball
production~\cite{Dimopoulos:2001qe} between the scales $M_s/g_s$
and $M_s/g_s^2$ at which point black holes start to be produced in
light of the BH string correspondence~\cite{bhstring}.  The cross
section for string ball production are interesting in and of
itself~\cite{Dimopoulos:2001qe}, but one only gets a parametric
separation of the black hole and string ball scales when looking
at a weakly coupled string theory, which is probably not the case
for RS in the IR.   With an $\mathcal{O}(1)$ string coupling, all
the scales would be about the same and one goes directly in BH
production above the scale $M$. In any case, string balls are
relevant only in the weakly coupled regime where black holes would
be out of reach.

We now consider string resonances. These resonances, even if too
wide to be seen explicitly in the cross section, will also affect
the angular distribution and lead to a rise at energies of order
the string scale. So probing the ratio $\reta$ can give a detailed
exploration of a weakly coupled string theory.

\begin{figure}[h]
\begin{center}
\includegraphics[width=8cm]{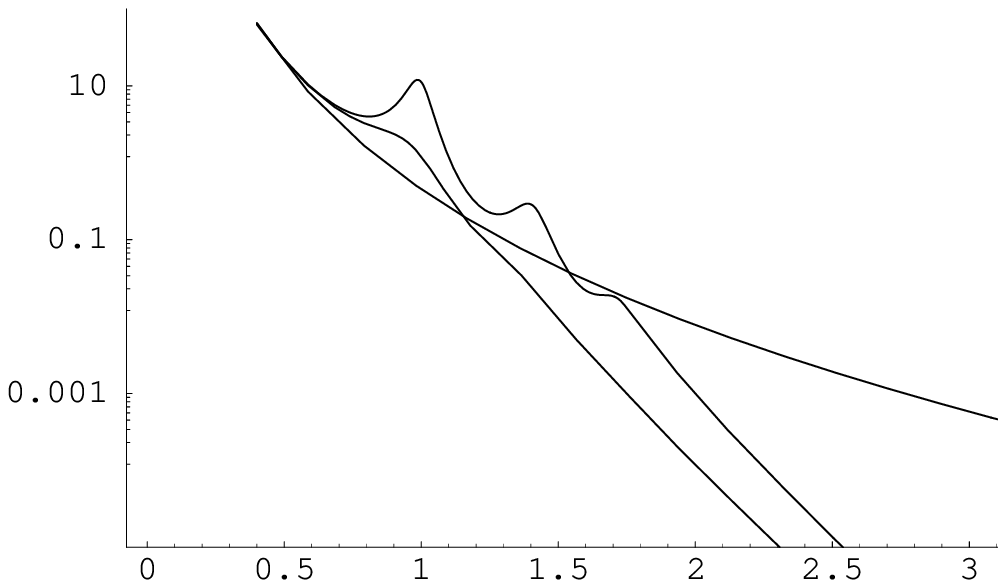}\includegraphics[width=8cm]{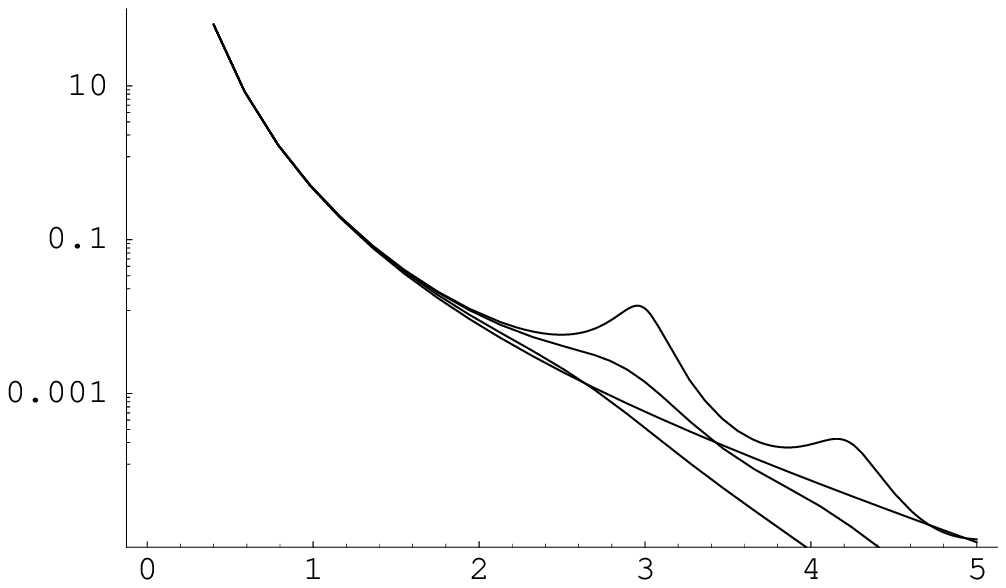}\\
\includegraphics[width=8cm]{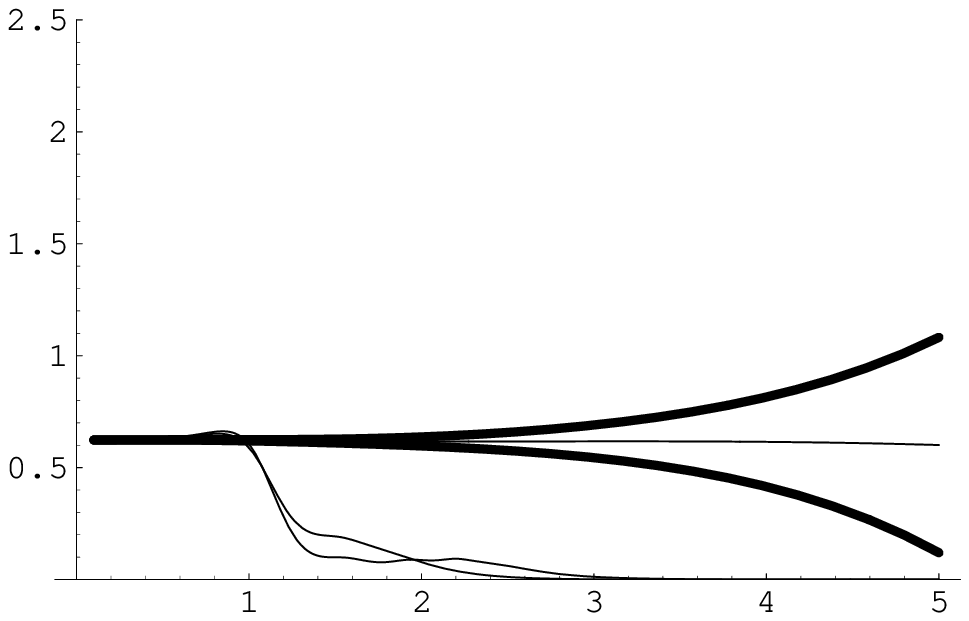}\includegraphics[width=8cm]{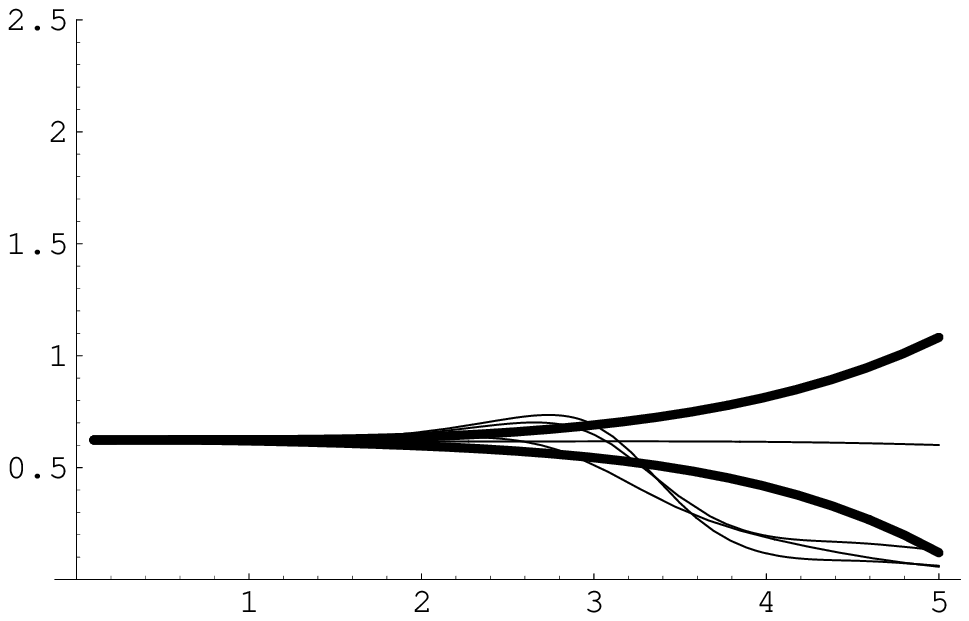}
\end{center}
\caption{In the upper plots $d\sigma/dM_{jj}$(units of pb/GeV) vs
$M_{jj} (TeV)$ is plotted for the case of SM QCD
background(thicker curve), and a toy stringy behavior with $M_s$=1
TeV in the lefthand plot with $\gamma=.1,.3$ and $M_s$=3 TeV in
the righthand plot with $\gamma=.1,.3,.6$.  In the lower two plots
the $\reta$ is plotted for the same parameters.}
\label{stringdsigdm}
\end{figure}

We now assume a model in which the Standard Model 2$\rightarrow$2
cross section is modified by a Veneziano-amplitude-motivated form factor:

  \begin{equation}
A_{pp\rightarrow jj}\equiv A_{SM} A_{ST}
\end{equation}
\begin{equation}\label{veznew}
A_{ST}\equiv \frac{\Gamma\left(1-\frac{s}{M_S^2}(1+i\gamma)\right)
\Gamma\left(1-\frac{t}{M_S^2}(1+i\gamma)\right)}{\Gamma\left(1-\frac{s}{M_S^2}(1+i\gamma)-\frac{t}{M_S^2}(1+i\gamma)\right)}
\end{equation}
yielding the results for the differential cross section and
$\reta$ shown in Figure~\ref{stringdsigdm}.

\begin{figure}[h]
\begin{center}
\includegraphics[width=8cm]{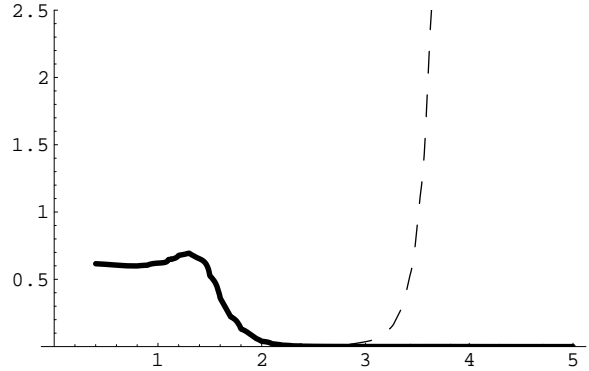}
\end{center}
\caption{$\reta$ is plotted for $M_s=1$ TeV and $\gamma=.1$ for
the amplitude defined with (\ref{vezorg}) with one sigma gaussian
error bars corresponding to 1 inverse femtobarn of luminosity.}
\label{stringerr}
\end{figure}

We see the characteristic string behavior. First of all we see
several resonances appear at about the string scale. Furthermore,
we see the 2$\rightarrow$2 cross section decrease in the region of
$\eta$ we have considered. But the most notable and characteristic
feature of string behavior would be the much less transverse
behavior of the 2$\rightarrow$2 cross section (exactly the
opposite of what we considered with black holes in the previous
subsection). We see this dramatically illustrated in the lower
plots, where $R$ goes from the QCD value of 0.6 (or higher when
there are resonances) down to essentially 0. In
Figure~\ref{stringerr} we have included statistical errors on
$\reta$ and we see that although $R$ is close to zero, there are
enough events to trust this value. That is, we can see the string
theory dip in the regime where there are sufficiently many events
to trust the result.

\begin{figure}[h]
\begin{center}
\includegraphics[width=8cm]{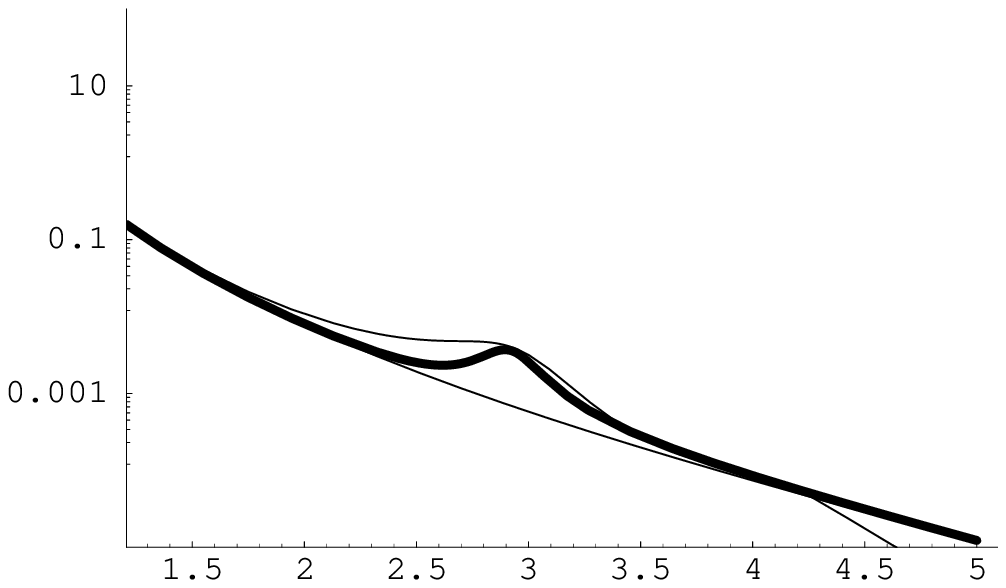}\includegraphics[width=8cm]{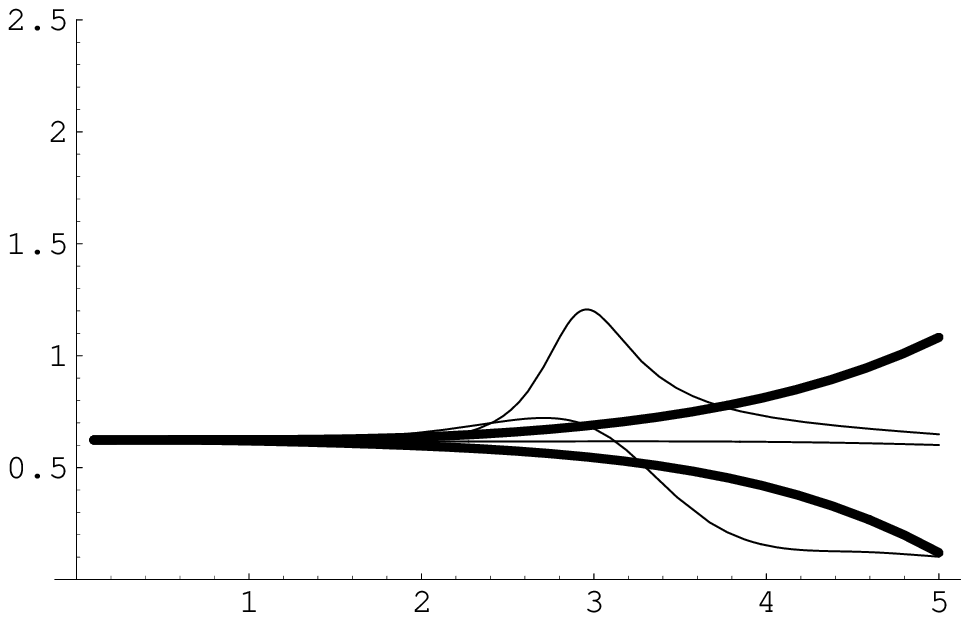}
\end{center}
\caption{In the left plot $d\sigma/dM_{jj}$(units of pb/GeV) vs
$M_{jj} (TeV)$ is plotted for the case of SM QCD background, a toy
stringy behavior with $M_s$=3 TeV and $\gamma=.2$ and a massive
colored octet resonance(thicker curve) with mass and width chosen
to mimic the differential cross section behavior near the
resonance.  In the right hand plot the same curves are plotted for
$\reta$, note the easily discernible difference between field
theory resonance and ``string" theory resonance.} \label{rescomp}
\end{figure}

It is also interesting to note that this dip allows us to readily
distinguish string theory from other models with resonances, such
as a colored octet resonance, as illustrated in
Figure~\ref{rescomp}. Even when $d\sigma/dM$ is rather similar,
$R$ reverts to the QCD value away from the resonance for the octet
but drops off for string behavior. In fact, the resonances might
not appear explicitly if they are too wide but we would still be
able to ascertain the presence of stringy physics.

\begin{figure}[h]
\begin{center}
\includegraphics[width=8cm]{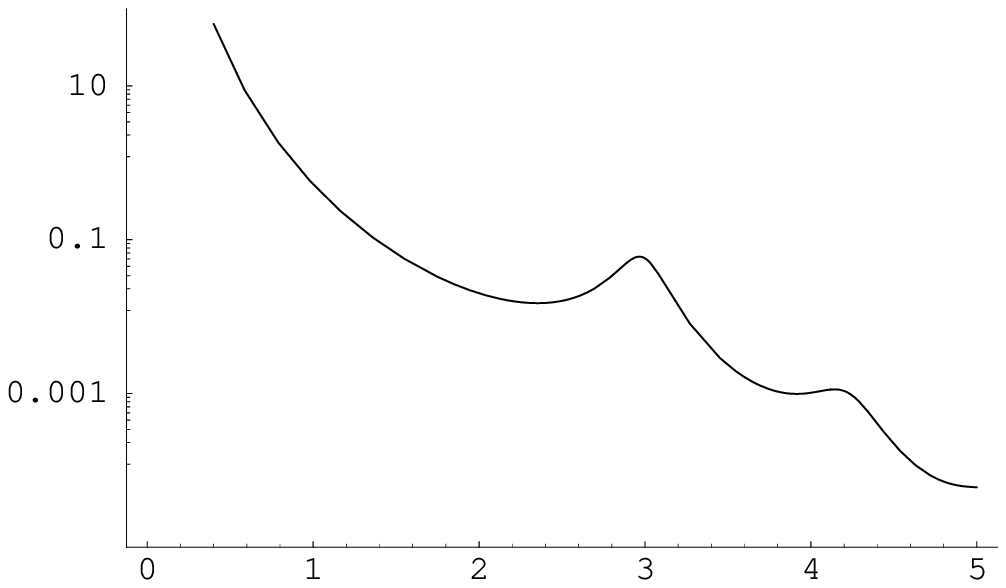}\includegraphics[width=8cm]{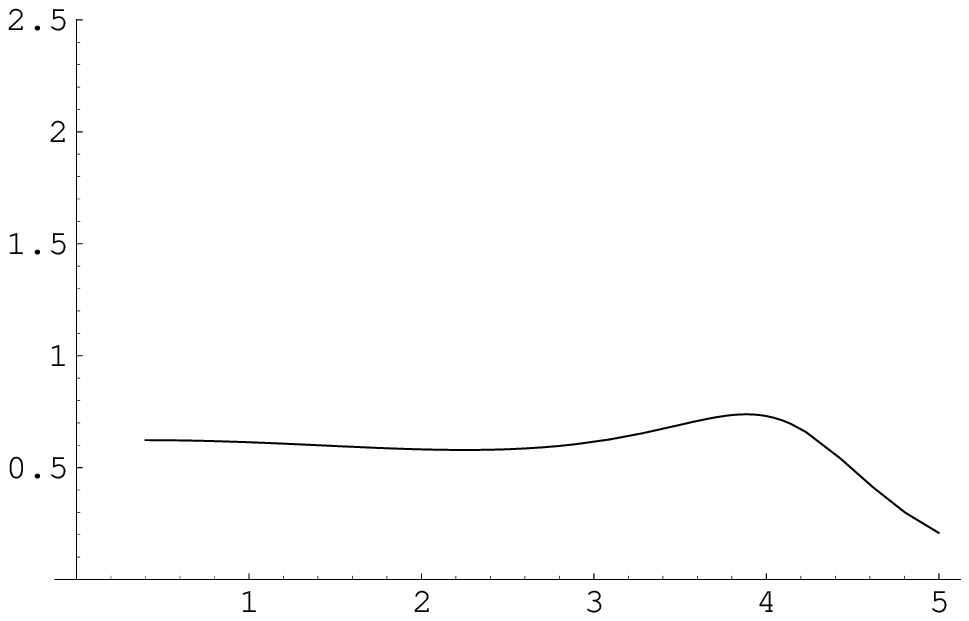}
\end{center}
\caption{In the left plot $d\sigma/dM_{jj}$(units of pb/GeV) vs
$M_{jj} (TeV)$ is plotted for the case of SM QCD
background(thicker curve), and a toy stringy
behavior~(\ref{vezorg}) with $M_s$=3 TeV and $\gamma=.1$.  In the
right hand plot $\reta$ is shown for the same parameters.}
\label{altstring}
\end{figure}

Finally, we show the somewhat notable result that with both the
differential cross section and the angular distribution we can
distinguish different stringy form factors.
\begin{equation}\label{vezorg}
A^0_{ST}\equiv
\frac{\Gamma\left(1-\frac{s}{M_S^2}(1+i\gamma)\right)
\Gamma\left(1-\frac{t}{M_S^2}(1+i\gamma)\right)}{\Gamma\left(2-\frac{s}{M_S^2}(1+i\gamma)-\frac{t}{M_S^2}(1+i\gamma)\right)}
\end{equation}
For example, with the original Veneziano amplitude (\ref{vezorg}),
there is no angular dependence in the first resonance.  Even
though there is a resonance appearing in the differential cross
section, it doesn't show up in $\reta$, as demonstrated in
Figure~\ref{altstring}. This would be a clean way to distinguish
the two different forms of the Veneziano amplitude, one which
appears in supersymmetric theories~(\ref{veznew}) and the original
amplitude~(\ref{vezorg}).

\subsubsection{Higher-Dimension Operators}\label{sec:higherdim}

Finally, strong gravity effects can give rise to higher dimension
operators that exist well below the gravity scale. The lowest
dimension operator would of course have dimension 6 but higher
dimension operators should also exist and be competitive when near
the gravity scale.  The effects of these operators can in some
ways resemble that of black holes-that is lead to a rise in cross
section above that of QCD and also give rise to much more
transverse events. However, there can be interesting features that
would distinguish these different contributions.

First is that the energy dependence of any particular
higher-dimension operator is distinctive, and can in principle
differentiate these operators from the "black hole" contribution
we described and from each other. As we will see when comparing he
higher dimensional results to the figures in
Section~\ref{sec:dibh}, this distinction will be most manifest in
the low energy region where the interference terms bring out the
energy dependence that is otherwise lost in the PDFs.

The second difference is that we assumed true black hole type
effects would have a threshold at about the quantum gravity scale
but have form factors that kill them at low energy.
Higher-dimension operators will by our assumption be those that
survive to low energy. Of course, it is possible that "black
holes" are responsible for higher-dimension operators that cut off
at low energy (or similarly grow rapidly above a certain scale as
in~\cite{Horava:1997dd}). In this case, they would be very much
like black hole type effects already considered and would
similarly cut off at low energies.

Although all higher-dimension operators might be relevant at the
LHC if the quantum gravity scale is low, we will focus on
four-fermion operators which are adequate to illustrate our point.
If the quantum gravity scale is high, it is appropriate to keep
the lowest dimension operator. If it is low, it would give a
qualitative sense of the behavior of the 2$\rightarrow$2 cross
section. The precise energy dependence can be quite different, but
this would  nonetheless look similar at higher
invariant mass where the operators would dominate over QCD so
interference terms would be insignificant. The current bound on
the scale suppressing a four fermion quark operator
$(q_L\gamma^\mu \bar{q}_L)^2$ where $q=u,d$ is $\Lambda=2.7$
TeV~\cite{pdg} assuming a coefficient of $2\pi$ for the operator.
CMS studies show that the mass scale that can be probed at the 95
\% C.L is $\Lambda \sim$ 15 TeV while at the 5 $\sigma$ discovery
level it can discover effects from $\Lambda\sim 12$ TeV with 10
inverse femtobarns of data.

In this regime one might also question the legitimacy of a higher
dimension operator parametrization. Over some of the energy regime
we are below the scale in the denominator. Of course the answer
depends on whether there are any small coupling factors etc. We
view this as a model. If the scales are comparable higher order
operators become relevant and eventually the expansion breaks down
altogether. But as the energy scale is not too far from the
denominator scale being probed, this shouldn't be too bad a model.

To find the form of the assumed four-fermion operators, we assume
that black holes respect gauge symmetries, but not necessarily
global symmetries. However, if we were to write down arbitrary
operators at the TeV scale generated by black holes then we would
immediately be ruled out by for instance proton decay and flavor
bounds.  Thus we  restrict the set of operators that we are
interested in to the lowest dimensional operators that preserve
the good symmetries of the SM.  This is of course a reasonable
assumption as we are inherently in the quantum gravity regime and
whatever theory this amounts to necessarily incorporates the
symmetries of the SM if the scale of quantum gravity is possibly
near the TeV scale.  This leaves us with dimension 6 operators of
the form
\begin{equation}
\sim \frac{c}{\Lambda^2}\Sigma (\bar{f}\gamma^\mu f)^2.
\end{equation}

Furthermore, the four fermion operator automatically accounts for
overall ``black hole" spin through its Lorentz structure whereas
in the black hole case one has different types of black holes with
different spins. One advantage of the four-fermion approach is
that it automatically takes into account the constraints of
spacetime symmetries. The four fermion operator automatically
produces the final states allowed by Lorentz symmetries
(symmetries which are violated by the classical black hole).

In addition to black holes with different angular momentum, black
holes could in principle carry different gauge charges. Again, the
four fermion operator and black hole would account for these in
different ways. In the four fermion case, it is a question of how
gauge indices are contracted. In general, for the black hole, we
assume it is charge neutral but this is not necessarily the case
since charged partons might collide to form the black hole.

As suggested above, the four-fermion operators might arise from
black holes or virtual black holes or some other nonperturbative
gravity effect. Higher-dimension operators might also arise from
perturbative loop calculations as demonstrated in Ref.
\cite{giudice} for example.

Four-fermion operators might also arise as virtual effects from
string theory at energies below the string scale. In fact, the
Veneziano amplitude (\ref{vezorg}) includes an operator that would
be accounted for by a four-fermion operator in its expansion (that
is an operator, with amplitude scaling as $s/M_S^2$. However, the
alternate form for the Veneziano amplitude that arises in
supersymmetric theories  (\ref{veznew}) does not have this term
and the first higher dimension operator enters at dimension 8. So
as we will see, exploring the energy dependence of the
differential cross section and $R$ could distinguish these
possibilities.  A particular example of modern string models with
dimension 6 operators can be found in~\cite{antoniadis} which can
be compared with the dimension 8 operators found in
~\cite{han1,peskin}.

Finally, string theory can give rise to harder scattering if in
warped space \cite{polchinskistrassler}. Whereas perturbative
string theory in flat space would cut off the 2$\rightarrow$2
scattering amplitude, perturbative string theory in warped space
is dual to a strongly interacting conformal field theory that
would give rise to hard scattering amplitudes. Ref.
\cite{polchinskistrassler} discusses how in curved space, hard
QCD-like behavior is reproduced in string theory, even though the
naive flat space expectation is that it does not. Moreover Bars
and Hinchliffe \cite{Bars:1985hh}, in their analysis of toy string
models for the SSC, have noted that if a string theory acted like
a QCD string then it would not demonstrate the dip in 2 body final
states characteristic of weakly coupled string theory in flat
space.  Additionally taking into non-perturbative string states as
in~\cite{Douglas:1996yp} could also produce harder scattering than
the naive flat space suppression of the string scattering cross
section, though for perturbative string theory this is a small
effect.

Of course the precise connection between the scale in the
denominator of the four-fermion operator and $M_D$ is model
dependent and in general unknown, as we will make even clearer in
the following subsection. Nonetheless, it is worthwhile looking
for such effects and distinguishing them from the other types of
quantum gravity behavior we have described.

\begin{figure}[h]
\begin{center}
\includegraphics[width=8cm]{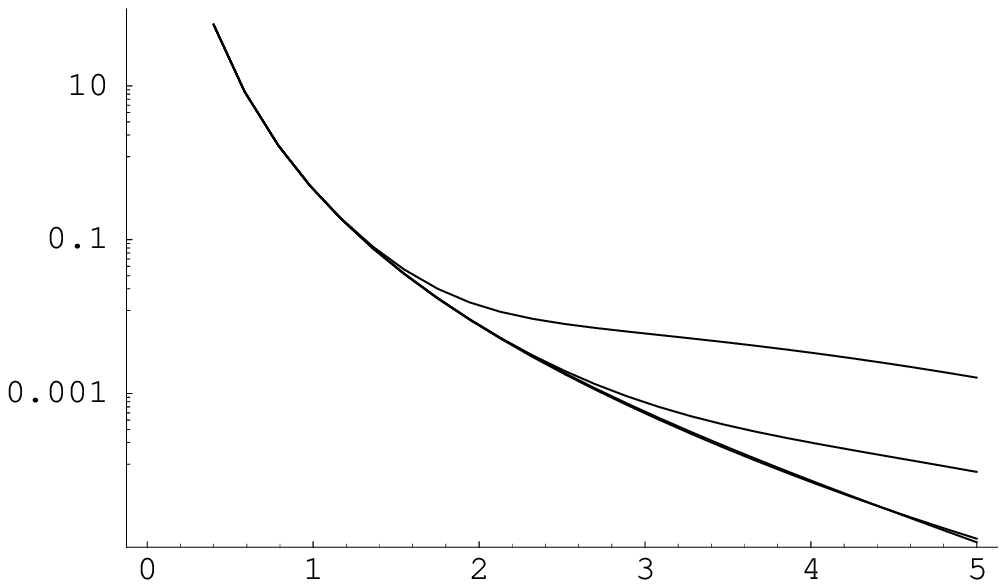}\includegraphics[width=8cm]{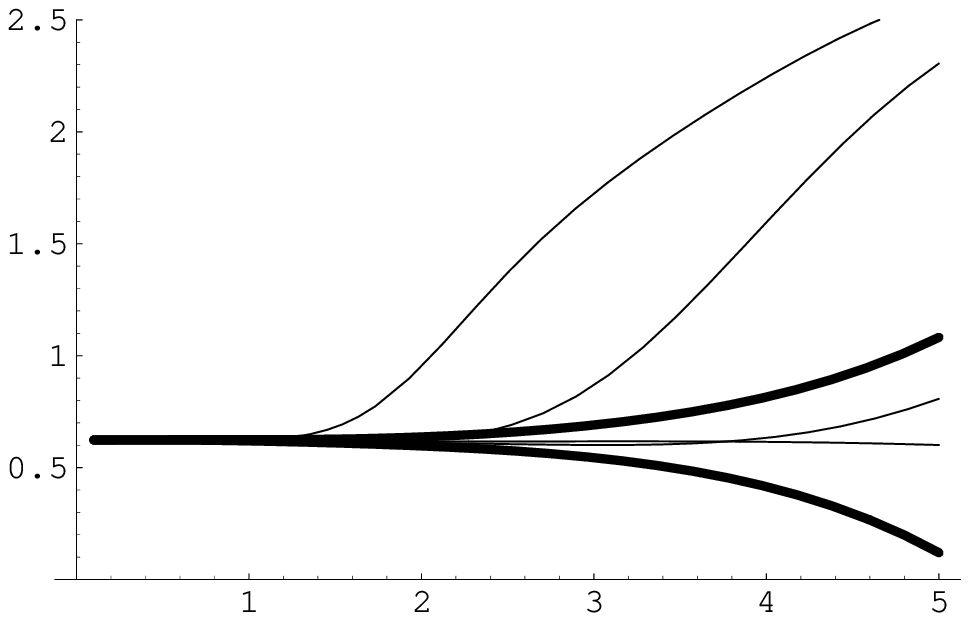}
\end{center}
\caption{In the left panel $d\sigma/dM$(units of pb/GeV) vs
$M_{jj}$(TeV) is plotted for QCD(the lowest curve) and a set of
four fermion operators with $\Lambda=1,2,4$ TeV.  In the right
panel $\reta$ is plotted for the same operators as well as QCD
with the statistical error bars for QCD are overlaid for 1
fb$^{-1}$ of data.} \label{fourferm}
\end{figure}

So we consider a four-fermion operator of the
form~\cite{Eichten:1983hw,Lane:1996gr}
\begin{equation}
\frac{1}{2\Lambda_{QG}^2} (\bar{\psi}_L\gamma^\mu\psi_L)^2
\end{equation}
Here we see the differential cross section and $\reta$ scale as in
Figure~\ref{fourferm}. Furthermore we can also potentially see
signals in the lepton channel as will be discussed in
Section~\ref{sec:lep}.

\subsection{Leptonic Final States}\label{sec:lep}

\begin{figure}[h]
\begin{center}
\includegraphics[width=8cm]{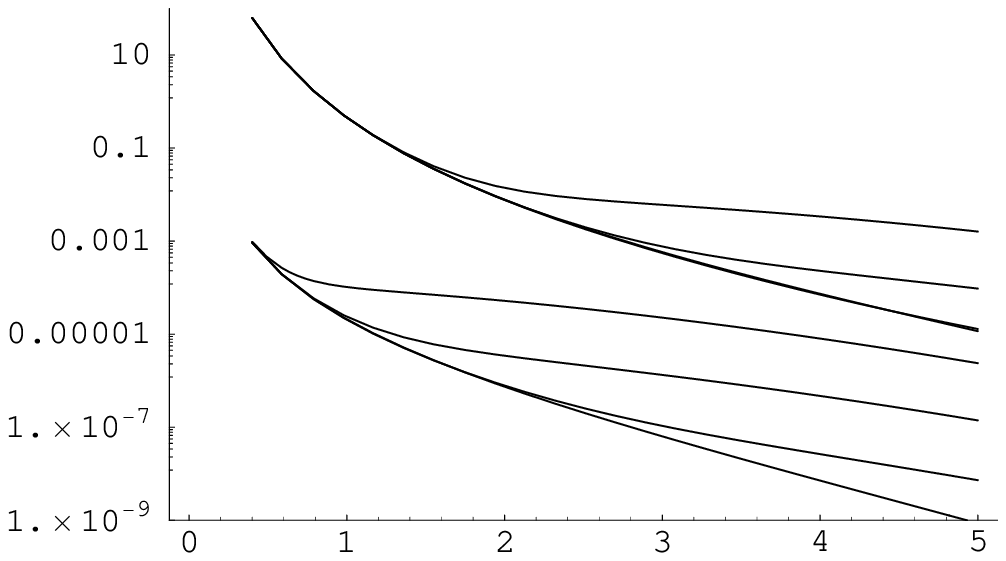}
\end{center}
\caption{ $d\sigma/dM$(units of pb/GeV) vs two body invariant
mass(TeV) is plotted for QCD(the lowest curve) and a set of four
fermion operators with $\Lambda=1,2,4$ TeV for dijets in the upper
curves.  In the lower curves SM Drell-Yan production of leptons is
plotted in combination with a four fermion operator that generates
a $l^{+}l^{-}$ final state with various $\Lambda=1,2,4$ TeV.}
\label{fourfermdy}
\end{figure}

  We have so far concentrated on jet final states which have the biggest
cross section. However, leptonic final states can be very
important as well and can be critical to distinguishing among
quantum gravity models. Leptons are generally clean enough at
these high energies that identifying new physics doesn't necessary
require studying the angular dependence. But for a black hole or
quantum gravity-related signature, we expect the same energy
dependence for the leptonic and hadronic final states.

  The most interesting feature of the leptonic final state will be the
relative cross sections for leptons and quarks. The relative ratio
would be different for a classical thermal decay (for which it is
about 10 \%) relative to the decay due to a four fermion operator,
for example, for which the relative rate is about 20 \%. In the
first case, the relative rates depend only on the number of
species and the multiplicity due to spin counting whereas in the
second case the numbers depend on number of species but also on
the dimension of the associated field. For example, gauge bosons
associated with a field strength would be suppressed relative to
fermions.  Although the first operators including a field strength
arise at dimension five $\bar{\psi}\psi\sigma_{\mu\nu}F^{\mu\nu}$
they are chirally suppressed (proportional to the light fermion
mass). Relative numbers of leptons vs quarks could also in
principle depend on whether the quarks and leptons are slightly
split from each other in a higher dimension, so a deviation from
naive predictions might indicate something about the structure of
the underlying theory.

But even more important than this counting is the way charge
flows. A black hole can in principle be formed from a color
charged initial state, namely two gluons or two quarks. The
leptonic fraction will then be 10 \% of the total black hole two
body decay rate.

The four fermion operators on the other hand would have to be
$q\bar{q}$ $l\bar{l}$ type operators which means that only the $q$
and $\bar{q}$ initial states will contribute. Because the PDFs for
$\bar{q}$ are so much smaller, the relative fraction of leptonic
final states will be much smaller. In addition, the four fermion
operator has an additional $\hat{u}/\hat{s}$ type suppression
which is relevant because we are focusing on the central region.
Finally a true four fermion operator will appear with an
additional alpha because the interference dominates. The upshot is
that the leptonic contribution from four-fermion operators is down
by about a factor of a thousand.

Nonetheless CMS studies show that the mass scale that can be
probed at the 5 $\sigma$ level can be as large as ~ 23 TeV for
dimuons~\cite{cmsdimuon} and 12 TeV in dijets~\cite{cmsdijet} for
10 inverse femtobarns (where the coefficient of the operator is
$2\pi$). The current strongest bounds from low energy experiments
are 4.2 TeV for dimuons and 2.7 TeV for dijets~\cite{pdg}. This
means that we might hope to distinguish various possibilities
through a joint measurement of lepton and jet final states.  In
Figure~\ref{fourfermdy} we plot the Drell-Yan background and the
effects of a $q\bar{q}$ $l\bar{l}$ operator as well as the dijet
operators from Section~\ref{sec:higherdim} to show a comparison of
rates for several values of $\Lambda$.

\begin{figure}[h]
\begin{center}
\includegraphics[width=8cm]{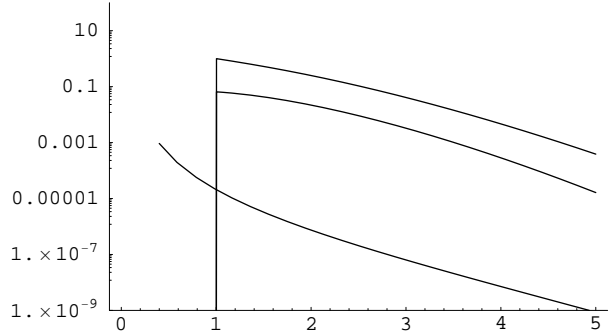}
\end{center}
\caption{ $d\sigma/dM$(units of pb/GeV) vs two body invariant
mass(TeV). The curves from top to bottom represent black hole
cross section for $M_D=1$ TeV , $n=6$, and $x_{min}=1$ for a black
hole decaying into $l^+l^-$ (assuming any initial gauge charge is
radiated softly), black hole cross section for $M_D=1$ TeV and
$x_{min}=1$ for a ``charged" black hole decaying into $l\nu$, and
the lowest curve is the Drell Yan background. } \label{dybhdj}
\end{figure}

Another possibility is that a black hole that will give rise to
the states had to be formed from a neutral state but didn't have
either the $\alpha$ or $\hat{u}/\hat{s}$ suppression factors.  It
is also possible that a charged final state is also of interest. A
$u$ and $\bar{d}$ initial state or the charge conjugate could give
rise to a charged state that can decay into lepton and neutrino.
We expect this final state to dominate above background as the
neutral state does.  In Figure \ref{dybhdj}, we plot these
possibilities and see the significant difference in overall rate
that can be used to distinguish among possibilities.

\section{Conclusions}

In this paper we have argued that if the higher-dimensional
gravity scale is indeed low, we are likely to learn more and
sooner about quantum gravity from studying two particle final
states than by studying multiparticle decays from
higher-dimensional black holes. Although we haven't precisely
determined the reach, we expect in all cases to be able to probe
to scales of order 5 TeV by looking at the two particle final
state channels.

We find it very unlikely that the LHC will produce conventional
black holes. In almost all cases the entropy is too low to trust
there is actually a black hole final state. In particular we found
even if we assume we are in the black hole regime that low
multiplicity final states dominate, and the two body final state
should be particularly interesting.

We have found a number of interesting features that can be used to
distinguish among the possibilities for quantum gravity effects.
We have shown that by studying both the differential cross section
and the transversality parameter $R_{\eta}$ as a function of
energy we can identify new effects and distinguish among black
hole type cross sections, perturbative string theory, and higher
dimension operators.

We have furthermore noted a number of specific points. The
relative lepton fraction can provide very valuable information. It
is likely to be large only in the case that black hole resonances
that shed gauge charge form. Otherwise the parton distribution
functions suppress the rate.

Furthermore we have found that we can distinguish string
resonances from other resonances and furthermore distinguish among
different string models.

We have found that black holes and four fermion operators differ
in their threshold behavior and that furthermore by studying the
threshold regime where interference is relevant one might be able
to distinguish different energy dependencies of various operators.

There are a number interesting follow up studies to consider.  On
the theory side, it would be nice to consider various string
models and their predictions. Although initial work on string
theory \cite{han1,han2,antoniadis,peskin} studied 2$\rightarrow$2
scattering, it would be interesting to see the predictions of
string models which have been developed in the interim.  It would
also be interesting to study threshold black hole behavior if at
all possible.

Other more phenomenological studies include seeing how much
information can be gleaned from the transversality or sphericity
of multibody final states. In particular, even if there is missing
energy or other jets, it might still be of interest to study the
two leading jets.

Finally a more detailed experimental analysis of how well one can
distinguish among different models would be very useful.

\section*{Acknowledgments}
We would like to thank Nima Arkani-Hamed, Tom Banks, Csaba
Cs\'{a}ki, Greg Landsberg, Lubos Motl, Alberto Nicolis, and Lenny
Susskind for useful discussions and Greg Landsberg for sharing his
results and a careful reading of our manuscript. The work of PM
and LR is supported in part by NSF grants PHY-0201124 and
PHY-0556111.

\appendix
\section{RS Black Holes}\label{app:rsbh}
\begin{figure}[h]
\begin{center}
\includegraphics[width=8cm]{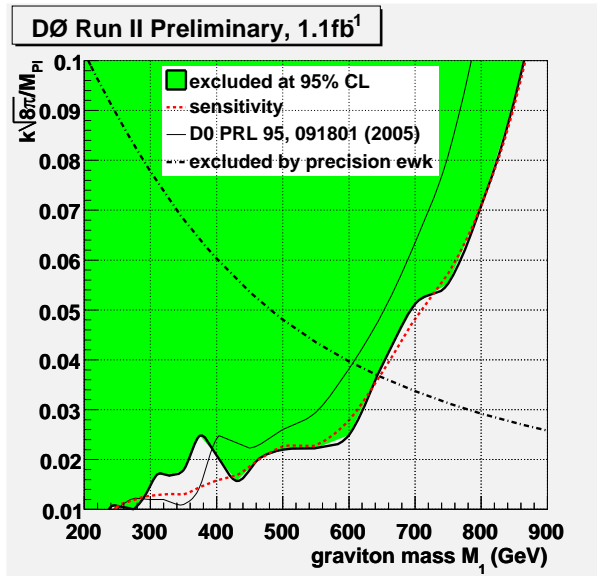}
\end{center}
\caption{Recent D0 search results from 1 inverse femtobarn of
luminosity updating the results of~\cite{Abazov:2005pi}.}
\label{d0results}
\end{figure}

We start with the convention for the RS1 normalization
\begin{equation}
\frac{M^3}{2}\int d^{5}x \sqrt{g} R
\end{equation}
which leads to the familiar relation
\begin{equation}
\mpl^2 = M^3/k(1-e^{-2kr_c})
\end{equation}
where $\mpl$ represents the reduced Planck mass, and $r_c$ is the
size of the extra dimension. To determine the relevant parameters
for black hole production in RS1 we need to reinterpret the
commonly used parameters $k/\mpl$ and $m_1$ (the first kk graviton
mass) in terms of the fundamental five dimensional Planck scale,
$M$, and the AdS curvature $k$.  The most current available bounds on
RS graviton production from the D0 experiment are given in
Figure~\ref{d0results}.  For a given value of $k/\mpl\equiv c$ and
$m_1$
\begin{equation}
\tilde{M}=\frac{m_1}{x_1 c^{2/3}},
\end{equation}
where $x_1=3.83$.  Thus if we were to extend $k/\mpl \sim .5$ and
take $1000$ GeV as the approximate bound for $m_1$ we get
$\tilde{M}\sim 350$ GeV.  Choosing $k/\mpl \sim .5$ means that the
strong coupling scale is extremely close to $k$ so to be somewhat
more conservative we use $\tilde{M}\sim 500$ GeV as the lower
bound that we have used throughout this paper when computing the
RS black hole cross section.

We now discuss what types of black holes exist in the context of
RS and their potential phenomenological implications. Black hole
solutions in Randall-Sundrum models have been studied for both RS1
and RS2 variants~\cite{stuff}. In studying RS black holes in
experiments the only relevant solutions are those in RS1 where
there is an IR brane with the SM localized there. The reason for
this is that in models without an IR brane the SM feels the usual
Planck scale, while in RS1 there is a warp factor that can allow
for an effective Planck scale of $\mathcal{O}(\mathrm{TeV})$ One
can consider  variants of this situation (see~\cite{rizzo}for
example) with light fermions and gluons in the bulk but this will
greatly suppress the black hole production cross section, since
they can be produced only at or near the TeV brane.

For RS1 black holes there are essentially two different regimes
that are relevant-- unlike flat space BHs.  In flat space one can
analyze black holes with Schwarzschild radius less than the
compactification radius and there is only one type of solution. In
RS1 since the bulk is warped with a scale set by the AdS curvature
$k$, there are separate regimes when $r_S << 1/(k e^{-kr_c})$ and
when $r_S \geq 1/(k e^{-kr_c})$ where curvature is relevant. In
the first case where $r_S << 1/(ke^{-kr_c})$ the black hole can be
thought of as a five dimensional flat space black hole, which
means the approximate expression for the Schwarzschild radius can
be obtained by matching the RS action to the Myers-Perry
solution\cite{myersperry} for a d-dimensional Schwarzschild BH.
Carrying out this matching one obtains
\begin{equation}\label{rsbh}
r_S=\left(\frac{M_{BH}}{3\pi^2 \tilde{M}^3}\right)^{1/2},
\end{equation}
where $\tilde{M}$ is the five dimensional Planck scale. Actually
it is subtle to derive this formula since the best way of deriving
it is coordinate dependent.

The simplest way to see that we expect five-dimensional RS almost
flat space black holes is to work in terms of the parameter
$\tilde{M}$ in the first place. Since we are interested only in
the region near the TeV brane, the warp factor doesn't even enter
(until we get to large distances) so we expect the behavior to be
that of five-dimensional gravity with a low Planck scale.

One can also  directly match
to the Myers-Perry solution using conformally flat coordinates as
in~\cite{Anchordoqui:2002fc}. In this case starting with the
metric
\begin{equation}\label{localcoord}
ds^2=\frac{1}{(kz)^2}\left(dz^2+dx_\mu^2\right)
\end{equation}
and performing a conformal transformation one is left with the
relevant part of the effective gravitational action at the TeV
brane
\begin{equation}
M^3 \left( \frac{1}{kz_0} \right)^3 \int d^4xdz
\sqrt{g}R=\tilde{M}^3\int d^4x dz\sqrt{g}R,
\end{equation}
where $z_0$ is the location of the TeV brane in the coordinates
(\ref{localcoord}), and $1/{z_0 k}$ is the warp factor.  From the
form of this solution it can be easily matched to (\ref{mpln}) the
Myers-Perry solution where the assumption is that the metric is
asymptotically flat.

In the commonly used RS coordinates
\begin{equation}\label{globcoord}
ds^2=\exp{-2ky}dx_\mu^2+dy^2
\end{equation}
the derivation isn't quite as obvious given that the effective
action at the TeV brane is
\begin{equation}
M^3 e^{-2kr_c} \int d^4x dy \sqrt{g} R.
\end{equation}
Thus there would seemingly not be enough powers of the warp factor
if we were matching to the Myers-Perry solution to end up with a
relevant black hole mass scale of $\tilde{M}$.  This should not
come as a surprise though as because in looking at fluctuations
about a given position in (\ref{globcoord}) the metric is not
manifestly 5D flat space.  However one can convince oneself that
the relevant mass scale for black hole production is $\tilde{M}$
by computing the effective gravitational potential at the TeV
brane,
\begin{equation}\label{gravpot}
V(r)\sim \frac{1}{\mpl^2} \frac{m_a m_b}{r}+\frac{1}{\mpl^2k
\exp{-3kr_c}} \frac{m_a m_b e^{-m_1 r}}{r^2}.
\end{equation}
Neglecting the first term which comes from the zero mode of the
graviton and is negligible at short distances, we see that the
second term approximates a five dimensional flat gravitational
potential for $r<1/m_1$.  From this one can derive the approximate
horizon radius via Laplace by setting the kinetic energy equal to
the potential energy (as also done for ADD black holes
in~\cite{Argyres:1998qn}) and thus we arrive at
\begin{equation}
r_S \sim \left(\frac{M_{BH}}{\tilde{M}^3}\right)^{1/2},
\end{equation}
where we have expanded the exponential in (\ref{gravpot}) to the
lowest order.  Additionally we see that from looking at higher
order terms in the expansion we should find deviations from this
approximate form at order $r_S\sim 1/k\exp{-kr_c}$ when we expect
curvature corrections to be taken into account.  Eventually when
the black holes are large enough in size the solution should
change to an AdS-Schwarzschild black hole.   From these various
derivations there are TeV sized approximate flat space black holes
in RS1 and we assume the flat space behavior for the relevant
regimes we are interested in throughout this paper.

As in flat space the production cross section for these black
holes is given roughly by $r_S^2$ and would seem to grow unbounded
with energy/the mass of the BH.  However in the context of RS
models we know this behavior has to be modified in some way as the
black hole size approaches the AdS curvature length. Additionally
one knows that the cross section cannot grow as a power law
forever from AdS/CFT reasoning~\cite{ahpr}\cite{}. In~\cite{ahpr}
it was conjectured that once you made black holes with size $1/k$
the cross section was bounded and never got larger than this
value. However a more refined understanding of what happens at
this scale was put forth by Giddings~\cite{giddingsfroissart}, who
conjectured from the gauge theory dual side at the scale $1/k$ the
Froissart bound is saturated and then the cross section doesn't
cease growing but grows as $\ln^2 E$. Interestingly enough
Giddings was also able to show that this behavior can be seen
directly by looking for BH solutions in linearized gravity and he
found that when taking into account curvature effects $r_S\sim \ln
E$ strengthening the argument for the Froissart bound $\ln^2 E$
behavior.

Given the additional effects of curvature in RS1 the basic regimes
for cross sections can be summarized in the following way
\begin{eqnarray}\label{diffxs}
E>\tilde{M}\;\;&&\sigma \sim E/\tilde{M}^3\nonumber \\
E\gappeq \left(\frac{M}{k}\right)^2 \tilde{M} \;\; &&\sigma \sim
\ln^2 E,
\end{eqnarray}
thus demonstrating that one gets rather different results than
simply a five dimensional flat space black hole depending on the
ratio $M/k$.  In practice however this effect is small when $M/k$
is large, as with QCD where   you never really see the effects of
the softening of the cross section).  We also see that we would
want $M/k$ to be large to approach the large entropy regime, but
this is the regime where experimental constraints are stronger.

This discussion of course ignores the possible effects of any
additional scales beyond $k$ and $M$, for instance an additional
scale set by $M_s$ and $g_s$. It is in principle possible to form
string resonances or string balls in this case as well as the
truly flat case.  Nevertheless as we see from (\ref{diffxs}) the
only really relevant parameter for RS1 ``black holes" is
$\tilde{M}$ given that black holes will dominantly be produced at
threshold due to the falling PDFs.

\section{Existing Constraints on the Quantum Gravity
Scale}\label{app:constraints}

In the text we have discussed several possible models for quantum
gravity effects, including black holes down to low energies,
perturbative string theory, and higher-dimensional operators.  In
all cases it will be difficult to constrain the Planck scale based
on nonobservation of these effects, in the first case because of
lack of knowledge of $x_{min}$, in the second case because we
don't know the string coupling, and in the latter case, because we
don't know how to predict a precise relationship between the
Planck scale and the scale occurring in a higher-dimensional
operator.

Nevertheless, we do need to consider other searches for the types
of operators and effects we have suggested, since in principle
they can rule them out over the measurable range. In particular,
Giudice, and Strumia as well as Contino et al
\cite{giudice,contino} considered four-fermion operators, while
\cite{antoniadis,han1,han2,peskin} considered constraints on the
string scale. In both cases, the constraints appear rather
stringent and if true, would significantly impinge on the
parameter regime we have considered.

We first consider Ref. \cite{giudice}, who demonstrate that loop
effects with KK gravitons exchanged can generate four-fermion
operators. They argued that the strong bounds on dimension-6
operators, in particular those involving quarks and leptons,
significantly constrain the allowed coefficients of the operator
they found. The operator is clearly suppressed by the
higher-dimensional Planck scale, $M^2$, and they chose to
interpret the experimental constraint as a constraint on the UV
cutoff on the effective theory for which a loop calculation would
apply. That is, string resonances or other states might  enter at
an energy lower then the strong scale determined by nave
dimensional analysis, and they put a constraint on the scale
$\Lambda$ for various values of $M_D$.

This analysis could have an impact on our study for several
reasons. First, if the cutoff necessarily occurs below the quantum
gravity scale, we would expect 2$\rightarrow$2 scattering to be
lower than the simple estimates we presented. Second,
independently of their cutoff procedure, if the scale of
four-fermion operators is constrained to be the current bound from
LEP II on quark lepton operators, the four fermion operators we
consider would already be ruled out.

We consider the second concern first. The current strongest bound
on the scale of four-fermion operator for $qqee$ with coefficient
$4 \pi/\Lambda^2$ is $\Lambda=26.4$ TeV~\cite{pdg}. However, there
are several reasons this quark lepton bound might not apply to
quantum gravity. The first is that the four quark operator and the
quark lepton operator will not necessarily occur at the same scale
if quarks and leptons are separated in the bulk, as they might be
to address baryon number violation concerns. A second reason this
operator might be suppressed is that if the operator is generated
by strong gravity, such as effects from black holes, the operator
might turn on only at high energy. Of course, this would not be a
true four fermion operator that applies to low energies but one
with significant form factor suppression at low energy. For the
true four fermion operator, if it is generated by strong gravity,
it could be that the lepton contribution is suppressed relative to
the quark contribution simply because there are fewer leptons
coupling to the graviton (and even fewer charged leptons). This
would not however prevent the loop contribution which is already
nominally too big for leptons. However it is possible that the
operator is not as stringent as presented in Ref. \cite{giudice}
because the cutoff for fields on the brane is different than than
in the bulk. We now reconsider this analysis.

The scale $\Lambda_S$ is  defined  in Ref. \cite{giudice} as the
strong scale  at which the gravitational coupling, $g^2=c_n
(\Lambda/M_D)^{2+n}$ (where $c_n$ in their convention for the
gravitational action is $(2\pi)^n$) is equal to the loop factor,
$(4 \pi)^{D/2} \Gamma(D/2)$, where $D=4+n$ is the number of
dimensions, and we will divide this loop factor into the product
$l_4 l_n$, where $l_4 = 16 \pi^2$.  So we have
\begin{equation}
\Lambda_{S}=\left( l_4 l_n \over c_n \right)^{1 \over 2+n} M_D
\end{equation}

If we were to compute the box diagram with two gravitons
exchanged,  with the graviton propagator, adding up all the KK
contributions, taken to be
\begin{equation}
G(k,0)={S_{n-1} \over 2 (2\pi)^n} \int^{\Lambda_{KK}^2}_0 dm^2
{(m^2)^{n/2-1} \over k^2 +m^2},
\end{equation}
where \begin{equation} S_{n-1}={2 \pi^{n/2} \over \Gamma(n/2)}.
\end{equation}
Yielding the box diagram contribution \cite{giudice}for the
coefficient of the four fermion operator
\begin{equation}
C_\gamma={15 \over 64} {c_n^2 \over l_4
M_D^{4+2n}}\int_0^{\Lambda^2}dk^2 k^4 G^2(k,0)\sim
\Lambda_S^{2+2n},
\end{equation}
where the scaling in the final expression comes from assuming all
momenta are cut off at $\Lambda_S$.

Notice that this answer would scale as $c_n^{2/(2+n)}/M_D^2$ as it
should for a convention independent answer.

If the authors of Ref. \cite{giudice} had cut off all momenta at
the scale $\Lambda_S$,  they would have concluded that over a
reasonable range of $M_D$, the strong scale determined by NDA is
already excluded. They chose instead to interpret this loop
contribution as implying the effective theory must be cut off at a
scale lower than $\Lambda_S$.

We will interpret the four-fermion operator bound differently and
assume the strong scale $\Lambda$ is  a function of the
convention-dependent parameter $M_D$ so we  interpret the bound
directly as a bound on the scale $M_D$. With this interpretation,
we would have a strong constraint on the allowed values of $M_D$
given the loop contribution that Giudice and Strumia computed.

However, in the phenomenological low scale gravity theory of
interest with respect to black hole production, the Standard Model
particles are confined to a brane whereas gravity  propagates
throughout the bulk (otherwise production is very suppressed but
for a scenario where black holes are investigated with matter in
the bulk see~\cite{rizzo}). The strong scale in Ref,
\cite{giudice} is based on the bulk particles and isn't
necessarily the cutoff for particles on the brane.

In fact,  with the cutoff taken to be the strong scale one notices
a peculiarity. Let us consider a box diagram with two gravitons
exchanged. As Giudice and Strumia point out, you get one integral
over four momentum suppressed by the phase space factor $l_4=16
\pi^2$ and you get two factors of $n$-dimensional momenta
suppressed by two factors of the $n$-dimensional phase space
factor $l_n$. If we assumed nothing cut off strong coupling, one
would find a loop diagram can generate a four fermion operator
with coefficient proportional to $1/l_4 \Lambda_S^{2+2n}$, which
is in turn proportional to $l_4^{n/(2+n)}$. That is, the answer
{\it grows} with the four-dimensional phase space factor, which is
very strange from the perspective of NDA. Generally NDA results
have phase space suppression in the denominator due to loop
integrals, some of which are partially compensated by the strong
scale, but never so as to grow with the phase factor.

The resolution to this puzzle presumably has to do with the fact
that we are considering dimension six four-fermion operators in
the first place. They are living only on the three plus one
dimensional surface of the brane. So although the integral is
higher dimensional, we expect that at least the additional
$n$-dimensional momentum integral should be cutoff in the
transverse directions by the size of the brane.  So  an
alternative NDA estimate is obtained by factoring the phase space
into the directions along the brane and the orthogonal directions.

One then finds the result scales as
\begin{equation}
\frac{15}{64}\frac{\pi^n c_n^2}{16\pi^2 (2\pi)^n \Gamma
(n/2)^2}\frac{\Lambda_S^2 \Lambda_{KK}^{2n}}{M_D^{4+2n}}
\end{equation}

The bound on the quantum gravity scale can be considerably relaxed
when the cutoff in the orthogonal directions is lower than
$\Lambda_S$ since $\Lambda_{KK}$ can be smaller than $\Lambda_S$.
For example, for $n=6$, $\Lambda_S=1.8 M_D$ whereas if
$\Lambda_{KK}=M/c_n^{1/(2+n)}$ (the convention-independent
quantity)  we have $\Lambda_{KK} =.25 M_D$. Since $\Lambda_{KK}$
is raised to the twelfth power, this gives a considerably smaller
result. Similarly, with $n=1$, we have $\Lambda_S=4.9 M_D$ and
$\Lambda_{KK}=0.54 M_D$. This is raised to the fourth power, again
considerably suppressing the final result. Of course
$\Lambda_{KK}$ can be bigger, but still satisfy the bound. For a
strong gravity theory, even without an explicit cutoff, we don't
know whether the gravity theory would actually apply to scales
$\Lambda_S$  which are bigger than $M_D$ in any case.

So we conclude that existing four fermion operator constraints are
serious but do not necessarily rule out the allowed parameter
space. It is therefore worthwhile to look for compositeness
effects of the type we described.

We now briefly turn to possible existing string bounds, which are
highly model-dependent. For example, Antoniadis \cite{antoniadis}
puts a bound on an operator scaling as $g_S/M_S^2$ where he takes
$g_S$ to be $g_{YM}^2=0.4$. With this assumption the string scale
is higher than about 3 TeV.

However, this bound assumes a particular model and a particular
string coupling. It could readily change by order unity if either
of these assumptions is abandoned. Furthermore, we don't know if
quark lepton operators are generated with the same or comparable
coefficients.

It should be noted that if we take weak string coupling to avoid
the bound, we don't improve the viability of black hole searches
since black holes would form at $M_S/g_s^2$ whereas string balls
would form at a scale $M_S/g_S$.

\end{document}